\documentclass{article}
\usepackage[utf8]{inputenc}
\usepackage{amsmath}
\usepackage{graphicx}
\usepackage{cite}

\title{Quaternion Mathematics in Electromagnetic Modeling and Simulation}
\author{Matthew David Marko\\matthew.marko.civ@us.navy.mil\\ \\Joe Schaff\\joe.schaff@gmail.com}
\date{22 September 2022}

\begin{document}

\maketitle

\section*{Abstract}
The purpose of this effort is to investigate if the use of quaternion mathematics can be used to better model and simulate the electromagnetic fields that occur from moving electromagnetic charges.  One observed deficiency with the commonly used Maxwell's equations is the issue of polar versus axial vectors; the electromagnetic field E is a polar vector, whereas the magnetic field B is an axial vector, where the direction of rotation remains the same even after the axial vector is inverted.  This effort first derived the rotation matrix for quaternion geometry.  This rotation matrix was then applied to modeling the magnetic fields at a distance from a source, and comparing it to traditional Maxwell's equations.  This effort was taken to model a series of moving charges, an observation aircraft observing a submarine, as well as an eddy current dynamic brake.  It was clearly observed that when a moving observer is studying a moving target, differences in the magnetic direction and magnitude can be observed, demonstrating the effectiveness of quaternion mathematics in such applications.  

\noindent NAVAIR Public Release 2022-450 Distribution Statement A - "Approved for public release; distribution is unlimited."

\section{Introduction}

The purpose of this effort is to investigate if the use of quaternion mathematics \cite{GA01_main, GA02, GA03, GA04, GA05, GA06, GA07, GA08} can be used to better model and simulate the electromagnetic fields that occur from moving electromagnetic charges.  One observed deficiency with the commonly used Maxwell's equations is the issue of polar versus axial vectors; the electromagnetic field E is a polar vector, whereas the magnetic field B is an axial vector, where the direction of rotation remains the same even after the axial vector is inverted.  It can be seen that the sense of rotation remains the same for the mirror image (axial vector), in contrast, the direction of motion of a particle, shown by the straight arrow, is inverted (polar vector). This illustrates the different natures of the electromagnetic field E (polar) and magnetic field B (axial) vectors.  

The quaternion-based geometric algebra model of electromagnetism developed and studied in this research provides better preservation of sign due to its non-commutative nature, and can also be represented by a single equation. The limitation of both this and the classical equations is that multiple charges or magnetic poles are treated as an aggregate for computation. This does not describe some of the subtle changes between the charges, or the differences between localized eddy currents, and their observed clustering. One way to add to the mathematical model / equations to provide better details of those electromagnetic phenomena is to incorporate a fractal aspect to the equations. Fractal structures have been observed in electromagnetic field distributions in both natural and synthetic environments.  Fractal patterns in synthetic electronic environments were first observed by Mandelbrot \cite{JS_R01} over 50 years ago in his research on transmission line noise, and in large volumes of packet data traffic in wide area network nodes by Paxson and Floyd \cite{JS_R02} over 25 years ago.

Magnetic anomalies have been observed in varying geophysical environments. The general diffusion equations through a geological medium is not sufficient to describe some coherent structures observed at multiple size scales. Moreover, the electromagnetic eddy currents and respective fields that traverse a geophysical medium have a fractal quality that has been observed in some media. 

These fractal aspects of EM fields in geophysical media \cite{JS_R03} are due to several factors that include: fractal diffusion layers which are caused in part by erosion of striated mineral deposits; existing fluids embodying fractional Brownian motion as described by Mandelbrot \cite{JS_R01}.  Random noise has been eliminated due to the observance of intrinsic structures. While this may describe most of the observed fractal structures frequently labeled as anomalous, there is still the fundamental observation by Voyager 2 of the magnetic fields at a significant distance of 25 AU from earth \cite{JS_R04}. The topology of these fields had a multifractal aspect to them, without the attributes needed to describe geophysical erosion, or similar changes to the medium. This was originally the motivation to try to rewrite the multipole (electric or magnetic) models using fractal dimensionality, in order to determine if such a model would have greater fidelity when compared to real-world observations.  

The discovery of fractal-based equations that better describe eddy currents and multipole charges will be the focus of future research. The groundwork for this has been started with this current research, specifically with the incorporation of a fractal noise model as a baseline for the current new equations. It is hoped that further investigations into the fractal nature of these real-world anomalies would provide a more comprehensive insight into anomalous EM field models.

Commonly, electromagnetic fields are represented with Maxwell's equation.  The most fundamental term is the electromagnetic field, commonly denoted as {\bf{E}}.  The force from a pair of electromagnetic charges is simply
\begin{eqnarray}
{F}_{12}&=&\frac{{Q_{1}}{\cdot}{Q_{2}}{\cdot}{{\bf{a}}_{12}}}/{{k_0}{\cdot}{R^2}},
\end{eqnarray}
where $Q_1$ and $Q_2$ are the charges (Coulombs), the normalized vector ${\bf{a}}_{12}$, the distance in between the two charges \emph{R} (m) squared, and $k_0$, where
\begin{eqnarray}
{k_0}&=&\frac{1}{4{\cdot}{\pi}{{\epsilon}_r}{\cdot}{{\epsilon}_0}}
\end{eqnarray}
where ${{\epsilon}_0}$ is the universal electric permittivity (${{10^{-9}/4{\cdot}{\pi}}}$), and ${{\epsilon}_r}$ is the relative permittivity of the material; where ${{\epsilon}_r}$ = 1 in a vacuum.  The electric field is simply the force for a given charge
\begin{eqnarray}
{\bf{E}}=\frac{\bf{F}}{Q_2}={Q_1}{\cdot}{\frac{\bf{a}_{12}}{k_0}}{\cdot}{R^2}. 
\end{eqnarray}
This is the electric field for a single point charge; the total electric field is the accumulation of all of the sources / fields.  

The magnetic field is a function of the electric field, and due to Faraday's law, the rate of change in the magnetic field is the curl of the electric field
\begin{eqnarray}
{\frac{{\partial}{\bf{B}}}{{\partial}{t}}}&=&-{\nabla}{\times}{\bf{E}}. 
\end{eqnarray}
Some parameters of the electric {\bf{E}} and magnetic {\bf{E}} field include that the gradient of the magnetic field is consistently 0 (Gauss' law of magnetism), 
\begin{eqnarray}
{\nabla}{\bullet}{\bf{B}}&=&0, 
\end{eqnarray}
the gradient of the electric field is proportional to the electric charge density $\rho$ (Gauss's law)
\begin{eqnarray}
{\nabla}{\bullet}{\bf{E}}&=&{\frac{\rho}{\epsilon_0}}, 
\end{eqnarray}
and the curl of the magnetic field is proportional to the current density {\bf{J}}, 
\begin{eqnarray}
{\nabla}{\times}{\bf{B}}-{{\frac{1}{c_0^2}}{\cdot}{\frac{{\partial}{\bf{E}}}{{\partial}t}}}&=&{\mu_0}{\cdot}{\bf{J}}, 
\end{eqnarray}
where ${\mu}_0$ is the magnetic permeability of a vacuum (${\mu}_0 = 4{\cdot}{\pi}{\cdot}{10^{-9}}$), and $c_0$ is the speed of light (${3{\cdot}10^8}$ m/s).  One relationship with the speed of light, the electric permittivity, and the magnetic permeability, 
\begin{eqnarray}
c_0&=&{\frac{1}{\sqrt{{{\mu}_0}{\cdot}{{\epsilon}_0}}}}.  
\end{eqnarray}

Geometric algebra is another approach, that maintains the distinction between axial and polar vectors.  Each of the three directions are treated as distinct algebraic imaginary variables ${e_1}$, ${e_2}$, and ${e_3}$.  When squared, they all equal a scalar
\begin{eqnarray}
{e_1^2}={e_2^2}={e_3^2}=1.
\end{eqnarray}
When the three of them are all multiplied together, they are equal to the imaginary value
\begin{eqnarray}
{e_1}{\cdot}{e_2}{\cdot}{e_3}={\sqrt{-1}}=j.
\end{eqnarray}
These variables anti-commute, where
\begin{eqnarray}
{e_1}{\cdot}{e_2}&=&-{e_1}{\cdot}{e_2},\\ \nonumber
{e_3}{\cdot}{e_1}&=&-{e_1}{\cdot}{e_3},\\ \nonumber
{e_3}{\cdot}{e_2}&=&-{e_2}{\cdot}{e_3}.\nonumber
\end{eqnarray}
Given the trivector quantity is the imaginary number, and that the vectors anti-commute, a dual relationship between the vectors and bivectors exist
\begin{eqnarray}
{e_1}{\cdot}{e_2}&=&{j}{\cdot}{e_3},\\ \nonumber
{e_3}{\cdot}{e_1}&=&{j}{\cdot}{e_2},\\ \nonumber
{e_2}{\cdot}{e_3}&=&{j}{\cdot}{e_1}.\nonumber
\end{eqnarray}
When using Geometric Algebra the imaginary bivector represents a rotation, as typically observed when comparing a magnetic field versus the vector electric field.  

Finally, when multiplying two vectors when using geometric algebra, it comes out to the gradient of the two vectors, plus the bivector quantity of the curl of the two vectors
\begin{eqnarray}
{\bf{vw}}&=&{{({{v_1}{\cdot}{e_1}}+{{v_2}{\cdot}{e_2}}+{{v_3}{\cdot}{e_3}})}{\cdot}{({{w_1}{\cdot}{e_1}}+{{w_2}{\cdot}{e_2}}+{{w_3}{\cdot}{e_3}})}}=\\ \nonumber
&=&{{v_1}{\cdot}{w_1}}+{{v_2}{\cdot}{w_2}}+{{v_3}{\cdot}{w_3}}+
{{({{v_1}{\cdot}{w_2}}-{{w_1}{\cdot}{v_2}})}{\cdot}{{e_1}{\cdot}{e_2}}}+
{{({{v_1}{\cdot}{w_3}}-{{w_1}{\cdot}{v_3}})}{\cdot}{{e_1}{\cdot}{e_3}}}+
{{({{v_2}{\cdot}{w_3}}-{{w_2}{\cdot}{v_3}})}{\cdot}{{e_2}{\cdot}{e_3}}}=\\ \nonumber
&=&{v{\bullet}w}+
{{({{v_1}{\cdot}{w_2}}-{{w_1}{\cdot}{v_2}})}{\cdot}{{j}{\cdot}{e_3}}}+
{{({{v_1}{\cdot}{w_3}}-{{w_1}{\cdot}{v_3}})}{\cdot}{{j}{\cdot}{e_2}}}+
{{({{v_2}{\cdot}{w_3}}-{{w_2}{\cdot}{v_3}})}{\cdot}{{j}{\cdot}{e_1}}}=\\ \nonumber
{\bf{vw}}&=&{v{\bullet}w}+{j{\cdot}{({\bf{v}}{\times}{\bf{w}})}}
\end{eqnarray}

When modeling electromagnetic fields using geometric algebra, all four Maxwell's equations can be combined into a single equation.  A common variable \emph{F} is used for geometric algebra of electromagnetic modeling
\begin{eqnarray}
F&=&{\bf{E}}+{j{\cdot}{c_0}{\cdot}{\bf{B}}}.
\end{eqnarray}
A singular equation for the electric field {\bf{E}}, the magnetic field {\bf{B}}, the current density {\bf{J}}, and the electric charge density $\rho$, can be written as
\begin{eqnarray}
{{[{{\frac{1}{c_0}}{\cdot}{(\frac{\partial}{{\partial}t})}}+{\nabla}]}F}&=&
{{\frac{\rho}{\epsilon_0}}-{{c_0}{\cdot}{\mu_0}{\cdot}{\bf{J}}}}.
\end{eqnarray}
This can be derived as
\begin{eqnarray}
{{[{{\frac{1}{c_0}}{\cdot}{(\frac{\partial}{{\partial}t})}}+{\nabla}]}F}&=&
{{[{{\frac{1}{c_0}}{\cdot}{(\frac{\partial}{{\partial}t})}}+{\nabla}]}{[{\bf{E}}+{j{\cdot}{c_0}{\cdot}{\bf{B}}}}]}={{\frac{1}{c_0}}{\cdot}{\frac{{\partial}{\bf{E}}}{{\partial}t}}}+{j{\cdot}{\frac{{\partial}{\bf{B}}}{{\partial}t}}}+{{\nabla}{\bf{E}}}+{j{\cdot}{c_0}{\cdot}{{\nabla}{\bf{B}}}}\nonumber \\ \nonumber
&=&{{\frac{1}{c_0}}{\cdot}{\frac{{\partial}{\bf{E}}}{{\partial}t}}}+{j{\cdot}{\frac{{\partial}{\bf{B}}}{{\partial}t}}}+{{\nabla}{\bullet}{\bf{E}}}+{j{\cdot}{\nabla}{\times}{\bf{E}}}+{j{\cdot}{c_0}{\cdot}{{\nabla}{\bullet}{\bf{B}}}-{{c_0}{\cdot}{{\nabla}{\times}{\bf{B}}}}}. 
\end{eqnarray}
Note the following fundamental electromagnetic equations
\begin{eqnarray}
{{\frac{1}{c_0}}{\cdot}{\frac{{\partial}E}{{\partial}t}}}&=&{{{c_0}{\cdot}{{\nabla}{\times}{\bf{B}}}-{{c_0}{\cdot}{\mu_0}{\cdot}{\bf{J}}}}}\nonumber \\ \nonumber
{\frac{{\partial}{\bf{B}}}{{\partial}t}}&=&{-{\nabla}{\times}{\bf{E}}}\\ \nonumber
{{\nabla}{\bullet}{\bf{B}}}&=&0
\end{eqnarray}
and therefore
\begin{eqnarray}
{{[{{\frac{1}{c_0}}{\cdot}{(\frac{\partial}{{\partial}t})}}+{\nabla}]}F}&=&
{{c_0}{\cdot}{{\nabla}{\times}{\bf{B}}}+{{c_0}{\cdot}{\mu_0}{\cdot}{\bf{J}}}}-{j{\cdot}{\nabla}{\times}{\bf{E}}}+{j{\cdot}{\nabla}{\times}{\bf{E}}}-{{c_0}{\cdot}{\mu_0}{\cdot}{\bf{J}}}\nonumber \\ \nonumber
&=&{{\frac{\rho}{\epsilon_0}}-{{c_0}{\cdot}{\mu_0}{\cdot}{\bf{J}}}}.
\end{eqnarray}

\section{Rotating the Reference Plane with Quaternion Geometry}

One of the advantages of quaternion mathematics for electromagetics is that during a rotation, the sense of rotation is preserved for axial vectors (such as the magnetic field), yet polar vectors (such as the electric field) get inverted.  For example, if inverting the reference frame entirely, where ${e_1}$ = -${e_1}$, ${e_2}$ = -${e_2}$, and ${e_3}$ = -${e_3}$, when dealing solely with Maxwell's equations, the electric field will invert as it should.  The magnetic field, however, ought to remain the same regardless of rotating the reference, yet the traditional Maxwell's equations will also erroneously invert the direction of the magnetic field.  With quaternions, the axial magnetic field vector maintains its direction.  Before inversion
\begin{eqnarray}
F&=&{\bf{E}}+{j{\cdot}{\bf{B}}},\nonumber \\
&=&{{E_1}{\cdot}{e_1}}+{{E_2}{\cdot}{e_2}}+{{E_3}{\cdot}{e_3}}+{j{\cdot}{({{B_1}{\cdot}{e_1}}+{{B_2}{\cdot}{e_2}}+{{B_3}{\cdot}{e_3}})}},\nonumber \\
&=&{{E_1}{\cdot}{e_1}}+{{E_2}{\cdot}{e_2}}+{{E_3}{\cdot}{e_3}}+{{{e_1}{\cdot}{e_2}{\cdot}{e_3}}{\cdot}{({{B_1}{\cdot}{e_1}}+{{B_2}{\cdot}{e_2}}+{{B_3}{\cdot}{e_3}})}},\nonumber 
\end{eqnarray}
If inverting \emph{F} to get \emph{F'},
\begin{eqnarray}
F&=&{{E_1}{\cdot}{-e_1}}+{{E_2}{\cdot}{-e_2}}+{{E_3}{\cdot}{-e_3}}+{{{-e_1}{\cdot}{-e_2}{\cdot}{-e_3}}{\cdot}{({{B_1}{\cdot}{-e_1}}+{{B_2}{\cdot}{-e_2}}+{{B_3}{\cdot}{-e_3}})}},\nonumber \\ 
&=&-({{E_1}{\cdot}{e_1}}+{{E_2}{\cdot}{e_2}}+{{E_3}{\cdot}{e_3}})+{{{e_1}{\cdot}{e_2}{\cdot}{e_3}}{\cdot}{({{B_1}{\cdot}{e_1}}+{{B_2}{\cdot}{e_2}}+{{B_3}{\cdot}{e_3}})}},\nonumber \\
&=&{-\bf{E}}+{j{\cdot}{\bf{B}}},\nonumber 
\end{eqnarray}
It is clear, by this inversion, that the electric field rotates with the rotation, and the magnetic field stays intact.  This rotation can be defined as a rotation matrix $\bar{R}$\\

$\bar{R}=
\begin{bmatrix}
-1 & 0 & 0\\
0 & -1 & 0\\
0 & 0 & -1\\
\end{bmatrix}\\$

\noindent It is observed that using quaternion geometric algebra, the magnetic field direction is preserved during the rotation, simply by rotating the “imaginary” magnetic field by the determinant of $\bar{R}$.  

The method of multiplying the rotation matrix by the determinant for quaternion geometry can be realized through the following derivation.  To understand how to best rotate a matrix with quaternion geometry, first the rotation matrix is identified: \\

$\begin{bmatrix}
{e_1'}\\
{e_2'}\\
{e_3'}\\
\end{bmatrix}$
=
$\begin{bmatrix}
{a_{11}} & {a_{21}} & {a_{31}}\\
{a_{12}} & {a_{22}} & {a_{32}}\\
{a_{13}} & {a_{23}} & {a_{33}}\\
\end{bmatrix}$
${\cdot}$
$\begin{bmatrix}
{e_1}\\
{e_2}\\
{e_3}\\
\end{bmatrix}\\$

\noindent which has the affect of: 
\begin{eqnarray}
{e_1'}&=&{{a_{11}}{\cdot}{e_1}}+{{a_{21}}{\cdot}{e_2}}+{{a_{31}}{\cdot}{e_3}}\nonumber \\
{e_2'}&=&{{a_{12}}{\cdot}{e_1}}+{{a_{22}}{\cdot}{e_2}}+{{a_{32}}{\cdot}{e_3}}\nonumber \\
{e_3'}&=&{{a_{13}}{\cdot}{e_1}}+{{a_{23}}{\cdot}{e_2}}+{{a_{33}}{\cdot}{e_3}}\nonumber
\end{eqnarray}

If one wants to rotate a magnetic field component during quaternion geometry, where 
$j = {{e_1}{\cdot}{e_2}{\cdot}{e_3}}$, this is resolved by rotating both and multiplying j with the vector.  The rotation of the imaginary component j is thus: 

\begin{eqnarray}
{e_1'}{\cdot}{e_2'}{\cdot}{e_3'}&=&(
{({{a_{11}}{\cdot}{e_1}}+{{a_{21}}{\cdot}{e_2}}+{{a_{31}}{\cdot}{e_3}})}{\cdot}
{({{a_{12}}{\cdot}{e_1}}+{{a_{22}}{\cdot}{e_2}}+{{a_{32}}{\cdot}{e_3}})}{\cdot}
{({{a_{13}}{\cdot}{e_1}}+{{a_{23}}{\cdot}{e_2}}+{{a_{33}}{\cdot}{e_3}})}\nonumber \\
&=&{({{a_{11}}{\cdot}{a_{12}}{\cdot}{e_1}{\cdot}{e_1}}+
{{a_{11}}{\cdot}{a_{22}}{\cdot}{e_1}{\cdot}{e_2}}+
{{a_{11}}{\cdot}{a_{32}}{\cdot}{e_1}{\cdot}{e_3}}+
{{a_{21}}{\cdot}{a_{12}}{\cdot}{e_2}{\cdot}{e_1}}+
{{a_{21}}{\cdot}{a_{22}}{\cdot}{e_2}{\cdot}{e_2}}}+\nonumber \\ \nonumber
&&{{a_{21}}{\cdot}{a_{32}}{\cdot}{e_2}{\cdot}{e_3}}+
{{a_{31}}{\cdot}{a_{12}}{\cdot}{e_3}{\cdot}{e_1}}+
{{a_{31}}{\cdot}{a_{22}}{\cdot}{e_3}{\cdot}{e_2}}+
{{a_{31}}{\cdot}{a_{32}}{\cdot}{e_3}{\cdot}{e_3}}){\cdot}
{({{a_{13}}{\cdot}{e_1}}+{{a_{23}}{\cdot}{e_2}}+{{a_{33}}{\cdot}{e_3}})}\\ \nonumber
&=&[({{a_{11}}{\cdot}{a_{12}}}+{{a_{21}}{\cdot}{a_{22}}}+{{a_{31}}{\cdot}{a_{32}}})+
{({{a_{11}}{\cdot}{a_{22}}}-{{a_{21}}{\cdot}{a_{12}}}){\cdot}{e_1}{\cdot}{e_2}}+
{({{a_{11}}{\cdot}{a_{32}}}-{{a_{31}}{\cdot}{a_{12}}}){\cdot}{e_1}{\cdot}{e_3}}+\\ \nonumber
&&{({{a_{21}}{\cdot}{a_{32}}}-{{a_{31}}{\cdot}{a_{22}}}){\cdot}{e_2}{\cdot}{e_3}}]{\cdot}
{({{a_{13}}{\cdot}{e_1}}+{{a_{23}}{\cdot}{e_2}}+{{a_{33}}{\cdot}{e_3}})},\\ \nonumber
&=&{({{a_{11}}{\cdot}{a_{12}}}+{{a_{21}}{\cdot}{a_{22}}}+{{a_{31}}{\cdot}{a_{32}}}){\cdot}{a_{13}}{\cdot}{e_1}}+
{({{a_{11}}{\cdot}{a_{12}}}+{{a_{21}}{\cdot}{a_{22}}}+{{a_{31}}{\cdot}{a_{32}}}){\cdot}{a_{23}}{\cdot}{e_2}}+\\ \nonumber &&
{({{a_{11}}{\cdot}{a_{12}}}+{{a_{21}}{\cdot}{a_{22}}}+{{a_{31}}{\cdot}{a_{32}}}){\cdot}{a_{33}}{\cdot}{e_3}}+
{{e_1}{\cdot}{e_2}{\cdot}{e_1}{\cdot}({a_{11}{\cdot}{a_{22}}}-{a_{21}{\cdot}{a_{12}}}){\cdot}{a_{13}}}+\\ \nonumber &&
{{e_1}{\cdot}{e_2}{\cdot}{e_3}{\cdot}({a_{11}{\cdot}{a_{22}}}-{a_{21}{\cdot}{a_{12}}}){\cdot}{a_{23}}}+
{{e_1}{\cdot}{e_3}{\cdot}{e_1}{\cdot}({a_{11}{\cdot}{a_{22}}}-{a_{21}{\cdot}{a_{12}}}){\cdot}{a_{33}}}+\\ \nonumber &&
{{e_1}{\cdot}{e_3}{\cdot}{e_1}{\cdot}({a_{11}{\cdot}{a_{32}}}-{a_{31}{\cdot}{a_{12}}}){\cdot}{a_{13}}}+
{{e_1}{\cdot}{e_3}{\cdot}{e_2}{\cdot}({a_{11}{\cdot}{a_{32}}}-{a_{31}{\cdot}{a_{12}}}){\cdot}{a_{23}}}+\\ \nonumber &&
{{e_1}{\cdot}{e_3}{\cdot}{e_3}{\cdot}({a_{11}{\cdot}{a_{32}}}-{a_{31}{\cdot}{a_{12}}}){\cdot}{a_{33}}}+
{{e_2}{\cdot}{e_3}{\cdot}{e_1}{\cdot}({a_{21}{\cdot}{a_{32}}}-{a_{31}{\cdot}{a_{22}}}){\cdot}{a_{13}}}+\\ \nonumber &&
{{e_2}{\cdot}{e_3}{\cdot}{e_2}{\cdot}({a_{21}{\cdot}{a_{32}}}-{a_{31}{\cdot}{a_{22}}}){\cdot}{a_{23}}}+
{{e_2}{\cdot}{e_3}{\cdot}{e_3}{\cdot}({a_{21}{\cdot}{a_{32}}}-{a_{31}{\cdot}{a_{22}}}){\cdot}{a_{33}}},
\end{eqnarray}

and therefore, 

\begin{eqnarray}
&=&{e_1}{\cdot}({{a_{11}}{\cdot}{a_{12}}{\cdot}{a_{13}}}+{{a_{21}}{\cdot}{a_{22}}{\cdot}{a_{13}}}+{{a_{31}}{\cdot}{a_{32}}{\cdot}{a_{13}}}+{{a_{11}}{\cdot}{a_{22}}{\cdot}{a_{23}}}
-{{a_{21}}{\cdot}{a_{12}}{\cdot}{a_{23}}}-{{a_{31}}{\cdot}{a_{12}}{\cdot}{a_{33}}}-{{a_{11}}{\cdot}{a_{32}}{\cdot}{a_{33}}})+\nonumber \\ \nonumber
&&{e_2}{\cdot}({{a_{11}}{\cdot}{a_{12}}{\cdot}{a_{23}}}+{{a_{21}}{\cdot}{a_{22}}{\cdot}{a_{23}}}+{{a_{31}}{\cdot}{a_{32}}{\cdot}{a_{23}}}+{{a_{21}}{\cdot}{a_{12}}{\cdot}{a_{13}}}
-{{a_{11}}{\cdot}{a_{22}}{\cdot}{a_{13}}}-{{a_{31}}{\cdot}{a_{22}}{\cdot}{a_{33}}}-{{a_{21}}{\cdot}{a_{32}}{\cdot}{a_{33}}})+\\ \nonumber
&&{e_3}{\cdot}({{a_{11}}{\cdot}{a_{12}}{\cdot}{a_{33}}}+{{a_{21}}{\cdot}{a_{22}}{\cdot}{a_{33}}}+{{a_{31}}{\cdot}{a_{32}}{\cdot}{a_{33}}}+{{a_{31}}{\cdot}{a_{12}}{\cdot}{a_{13}}}
-{{a_{11}}{\cdot}{a_{32}}{\cdot}{a_{13}}}-{{a_{31}}{\cdot}{a_{22}}{\cdot}{a_{23}}}-{{a_{21}}{\cdot}{a_{32}}{\cdot}{a_{23}}})+\\ \nonumber
&&{e_1}{\cdot}{e_2}{\cdot}{e_3}{\cdot}({{a_{11}}{\cdot}{a_{22}}{\cdot}{a_{33}}}-{{a_{21}}{\cdot}{a_{12}}{\cdot}{a_{33}}}+{{a_{31}}{\cdot}{a_{12}}{\cdot}{a_{23}}}-
{{a_{11}}{\cdot}{a_{32}}{\cdot}{a_{23}}}+{{a_{21}}{\cdot}{a_{32}}{\cdot}{a_{13}}}-{{a_{31}}{\cdot}{a_{22}}{\cdot}{a_{13}}}).  
\end{eqnarray}
As a result, the rotation matrix of an imaginary value is rotated with the following relationship
\begin{eqnarray}
{e_1'}{\cdot}{e_2'}{\cdot}{e_3'}&=&{{\bar{E}_1}{\cdot}{e_1}}+{{\bar{E}_2}{\cdot}{e_2}}+{{\bar{E}_3}{\cdot}{e_3}}+{{\bar{E}_x}{\cdot}{e_1}{\cdot}{e_2}{\cdot}{e_3}}\nonumber \\
&=&{{\bar{E}_1}{\cdot}{e_1}}+{{\bar{E}_2}{\cdot}{e_2}}+{{\bar{E}_3}{\cdot}{e_3}}+{j{\cdot}{\bar{E}_x}},\nonumber
\end{eqnarray}
where
\begin{eqnarray}
{\bar{E}_1}&=&({{a_{11}}{\cdot}{a_{12}}{\cdot}{a_{13}}}+{{a_{21}}{\cdot}{a_{22}}{\cdot}{a_{13}}}+{{a_{31}}{\cdot}{a_{32}}{\cdot}{a_{13}}}+{{a_{11}}{\cdot}{a_{22}}{\cdot}{a_{23}}}
-{{a_{21}}{\cdot}{a_{12}}{\cdot}{a_{23}}}-{{a_{31}}{\cdot}{a_{12}}{\cdot}{a_{33}}}-{{a_{11}}{\cdot}{a_{32}}{\cdot}{a_{33}}})\nonumber \\
{\bar{E}_2}&=&({{a_{11}}{\cdot}{a_{12}}{\cdot}{a_{23}}}+{{a_{21}}{\cdot}{a_{22}}{\cdot}{a_{23}}}+{{a_{31}}{\cdot}{a_{32}}{\cdot}{a_{23}}}+{{a_{21}}{\cdot}{a_{12}}{\cdot}{a_{13}}}
-{{a_{11}}{\cdot}{a_{22}}{\cdot}{a_{13}}}-{{a_{31}}{\cdot}{a_{22}}{\cdot}{a_{33}}}-{{a_{21}}{\cdot}{a_{32}}{\cdot}{a_{33}}})\nonumber \\
{\bar{E}_3}&=&({{a_{11}}{\cdot}{a_{12}}{\cdot}{a_{33}}}+{{a_{21}}{\cdot}{a_{22}}{\cdot}{a_{33}}}+{{a_{31}}{\cdot}{a_{32}}{\cdot}{a_{33}}}+{{a_{31}}{\cdot}{a_{12}}{\cdot}{a_{13}}}
-{{a_{11}}{\cdot}{a_{32}}{\cdot}{a_{13}}}-{{a_{31}}{\cdot}{a_{22}}{\cdot}{a_{23}}}-{{a_{21}}{\cdot}{a_{32}}{\cdot}{a_{23}}})\nonumber \\
{\bar{E}_X}&=&({{a_{11}}{\cdot}{a_{22}}{\cdot}{a_{33}}}-{{a_{21}}{\cdot}{a_{12}}{\cdot}{a_{33}}}+{{a_{31}}{\cdot}{a_{12}}{\cdot}{a_{23}}}-
{{a_{11}}{\cdot}{a_{32}}{\cdot}{a_{23}}}+{{a_{21}}{\cdot}{a_{32}}{\cdot}{a_{13}}}-{{a_{31}}{\cdot}{a_{22}}{\cdot}{a_{13}}}).\nonumber
\end{eqnarray}

When the rotation of a magnetic field $j{\cdot}{c_0}{\cdot}B$ occurs with quaternion geometry, \\ \\
$\begin{bmatrix}
{e_1'}\\
{e_2'}\\
{e_3'}\\
\end{bmatrix}$
=
$\begin{bmatrix}
{a_{11}} & {a_{21}} & {a_{31}}\\
{a_{12}} & {a_{22}} & {a_{32}}\\
{a_{13}} & {a_{23}} & {a_{33}}\\
\end{bmatrix}$
${\cdot}$
$\begin{bmatrix}
{j{\cdot}{e_1}}\\
{j{\cdot}{e_2}}\\
{j{\cdot}{e_3}}\\
\end{bmatrix}$
=

$\begin{bmatrix}
({{a_{11}{\cdot}{e_1}}+{{a_{21}{\cdot}{e_2}}+{{a_{31}{\cdot}{e_3}}){\cdot}({{\bar{E}_1}{\cdot}{e_1}}+{{\bar{E}_2}{\cdot}{e_2}}+{{\bar{E}_3}{\cdot}{e_3}}+{j{\cdot}{\bar{E}_x}}}}})\\
({{a_{12}{\cdot}{e_1}}+{{a_{22}{\cdot}{e_2}}+{{a_{32}{\cdot}{e_3}}){\cdot}({{\bar{E}_1}{\cdot}{e_1}}+{{\bar{E}_2}{\cdot}{e_2}}+{{\bar{E}_3}{\cdot}{e_3}}+{j{\cdot}{\bar{E}_x}}}}})\\
({{a_{13}{\cdot}{e_1}}+{{a_{23}{\cdot}{e_2}}+{{a_{33}{\cdot}{e_3}}){\cdot}({{\bar{E}_1}{\cdot}{e_1}}+{{\bar{E}_2}{\cdot}{e_2}}+{{\bar{E}_3}{\cdot}{e_3}}+{j{\cdot}{\bar{E}_x}}}}})\\
\end{bmatrix}$
=
$\begin{bmatrix}
{Q_1}\\
{Q_2}\\
{Q_3}\\
\end{bmatrix}$

\begin{eqnarray}
Q_{i}&=&
{{E_1}{\cdot}{a_{1i}}{\cdot}{e_1}{\cdot}{e_1}}+
{{E_1}{\cdot}{a_{2i}}{\cdot}{e_1}{\cdot}{e_2}}+
{{E_1}{\cdot}{a_{3i}}{\cdot}{e_1}{\cdot}{e_3}}+
{{E_2}{\cdot}{a_{1i}}{\cdot}{e_2}{\cdot}{e_1}}+
{{E_2}{\cdot}{a_{2i}}{\cdot}{e_2}{\cdot}{e_2}}+
{{E_2}{\cdot}{a_{3i}}{\cdot}{e_2}{\cdot}{e_3}}+\nonumber \\ \nonumber &&
{{E_3}{\cdot}{a_{1i}}{\cdot}{e_3}{\cdot}{e_1}}+
{{E_3}{\cdot}{a_{2i}}{\cdot}{e_3}{\cdot}{e_2}}+
{{E_3}{\cdot}{a_{3i}}{\cdot}{e_3}{\cdot}{e_3}}+
{{E_X}{\cdot}{a_{1i}}{\cdot}{j}{\cdot}{e_1}}+
{{E_X}{\cdot}{a_{2i}}{\cdot}{j}{\cdot}{e_2}}+
{{E_X}{\cdot}{a_{3i}}{\cdot}{j}{\cdot}{e_3}}.\\
&=&{{E_1}{\cdot}{a_{1i}}}
+{{E_1}{\cdot}{a_{2i}}{\cdot}{j}{\cdot}{e_3}}
-{{E_1}{\cdot}{a_{3i}}{\cdot}{j}{\cdot}{e_2}}
-{{E_2}{\cdot}{a_{1i}}{\cdot}{j}{\cdot}{e_3}}
+{{E_2}{\cdot}{a_{2i}}}
+{{E_2}{\cdot}{a_{3i}}{\cdot}{j}{\cdot}{e_1}}+\nonumber \\ \nonumber &&
{{E_3}{\cdot}{a_{1i}}{\cdot}{j}{\cdot}{e_2}}
-{{E_3}{\cdot}{a_{2i}}{\cdot}{j}{\cdot}{e_1}}
+{{E_3}{\cdot}{a_{3i}}}
+{{E_X}{\cdot}{a_{1i}}{\cdot}{j}{\cdot}{e_1}}
+{{E_X}{\cdot}{a_{2i}}{\cdot}{j}{\cdot}{e_2}}
+{{E_X}{\cdot}{a_{3i}}{\cdot}{j}{\cdot}{e_3}}\\ \nonumber
&=&({{E_1}{\cdot}{a_{1i}}}+{{E_2}{\cdot}{a_{2i}}}+{{E_3}{\cdot}{a_{3i}}})+
({{E_2}{\cdot}{a_{3i}}}-{{E_3}{\cdot}{a_{2i}}}+{{E_X}{\cdot}{a_{1i}}}){\cdot}{j}{\cdot}{e_1}+\\ \nonumber &&
(-{{E_1}{\cdot}{a_{3i}}}+{{E_3}{\cdot}{a_{1i}}}+{{E_X}{\cdot}{a_{2i}}}){\cdot}{j}{\cdot}{e_2}+
({{E_1}{\cdot}{a_{2i}}}-{{E_2}{\cdot}{a_{1i}}}+{{E_X}{\cdot}{a_{3i}}}){\cdot}{j}{\cdot}{e_3}
\end{eqnarray}
\noindent It was observed that a much simpler approach to rotating the magnetic field is to multiply the rotation matrix by the determinant of the rotation matrix when rotating the magnetic field; this is not an exact simplification but often a good approximation.  If rotating in units of 90$^{\circ}$, it matches exactly.  

\section{Model Development}

\begin{figure}
\centering
\includegraphics[width=\textwidth]{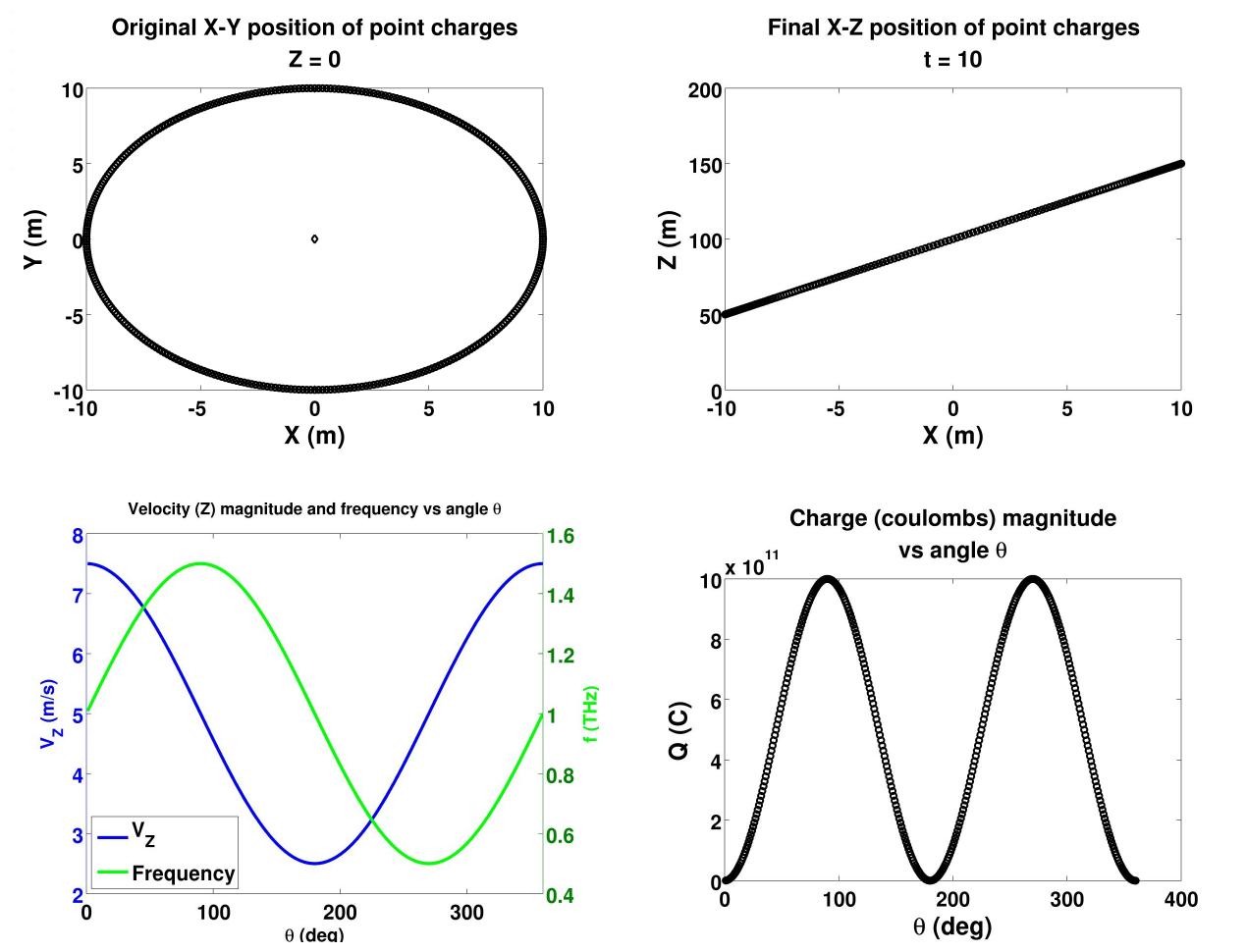}\\
\caption{Point-charge simulation parameters.}\label{fig:fig01}
\end{figure}

The model (Figure \ref{fig:fig01}) uses 360 point charges, spread in a circle in the X-Y plane, 10 meters apart from the reference point.  The charge varied with a sinusoidal function
\begin{eqnarray}
Q&=&{10^{12}}{\cdot}{sin(2{\cdot}{\theta})}, 
\end{eqnarray}
where $0<{\theta}<2{\cdot}{\pi}$.  The velocity is only in the Z-direction, going in both directions, at a velocity $V_Z$ (m/s) of 
\begin{eqnarray}
V_Z&=&5{\cdot}{cos({\theta})}{\cdot}[1+{0.5{\cdot}cos(2{\cdot}{\pi}{\cdot}f{\cdot}t)}], 
\end{eqnarray}
where \emph{t} (s) is the time, and \emph{f} (Hz) is the frequency of the velocity fluctuation, defined as
\begin{eqnarray}
f={10^{12}}{\cdot}(1+0.5{\cdot}{sin(\theta)}). 
\end{eqnarray}
The simulation will run for 10 seconds, discretized into 1,000 time steps of 0.01 second.  

\begin{figure}
\centering
\includegraphics[width=\textwidth]{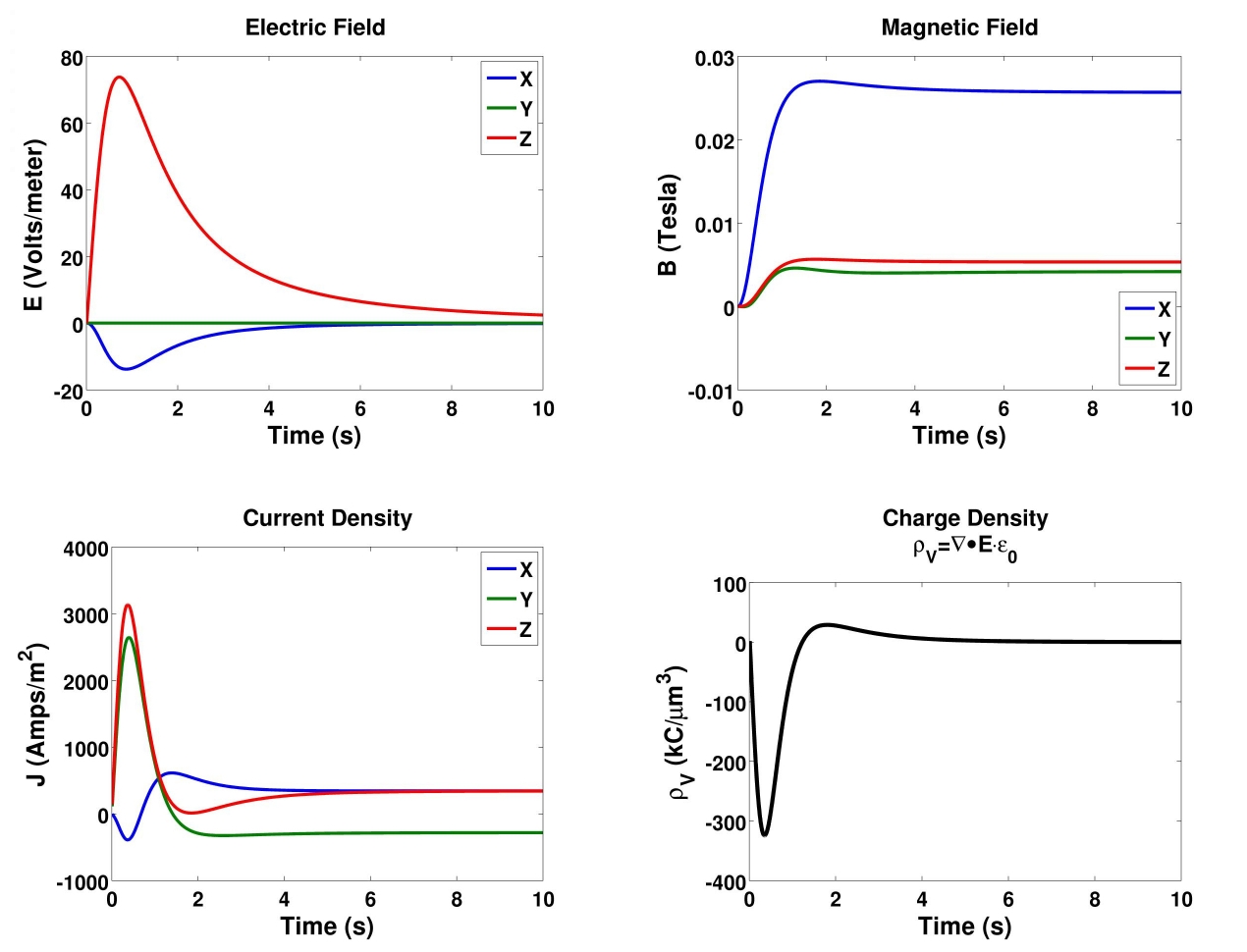}\\
\caption{Point-charge results (Maxwell's).}\label{fig:fig02}
\end{figure}

Before focusing on quaternion's, the simulation solved the Maxwell's equations for the electric field {\bf{E}} (Volts / meter), the magnetic field {\bf{B}} (Tesla), the current density {\bf{J}} (Amps/m${^2}$), and the charge volumetric density $\rho$ (C/m${^3}$).  The model worked by calculating the contribution of each charge to the electric field both at the null reference point (0,0,0), but also in a 5$\times$5$\times$5 3-D grid of positions, separated by a small distance increment of ${\delta}x$ = 1 cm.  This distance increment is used to find the change of electric field at the reference point, both at the reference point
\begin{eqnarray}
{\frac{{\partial}{E_1}}{{\partial}x}}{(0,0,0)}&=&{\frac{{E_1({\delta}x,0,0)}-{E_1{(-{\delta}x,0,0)}}}{2{\cdot}{\delta}x}},\nonumber
\end{eqnarray}
and at the neighboring points
\begin{eqnarray}
{\frac{{\partial}{E_1}}{{\partial}x}}{({{\delta}x},0,0)}&=&{\frac{{E_1(2{\cdot}{\delta}x,0,0)}-{E_1{(-{\delta}x,0,0)}}}{2{\cdot}{\delta}x}}.\nonumber
\end{eqnarray}
The differential of the electric field at the reference point is necessary to find the magnetic field {\bf{B}} at the reference point due to Faraday's Law
\begin{eqnarray}
{\frac{{\partial}{\bf{B}}}{{\partial}t}}=-{\nabla}{\times}{\bf{E}}.\nonumber
\end{eqnarray}
The electric field differential at the neighboring grid points is used to find the magnetic field at the neighboring grid points, so that the differential of the magnetic field at the reference point can be determined, so that the curl of the magnetic field can be used to find the current density {\bf{J}} due to Ampère's Law
\begin{eqnarray}
{{\nabla}{\times}{\bf{B}}-{{\frac{1}{c_0^2}}{\cdot}{\frac{{\partial}{\bf{E}}}{{\partial}t}}}}&=&{{{\mu}_0}{\cdot}{\bf{J}}}.
\end{eqnarray}
The results of this study are plotted in Figure \ref{fig:fig02}.  

\subsection{Parametric Study Initial Conditions}

\begin{figure}
\centering
\includegraphics[width=\textwidth]{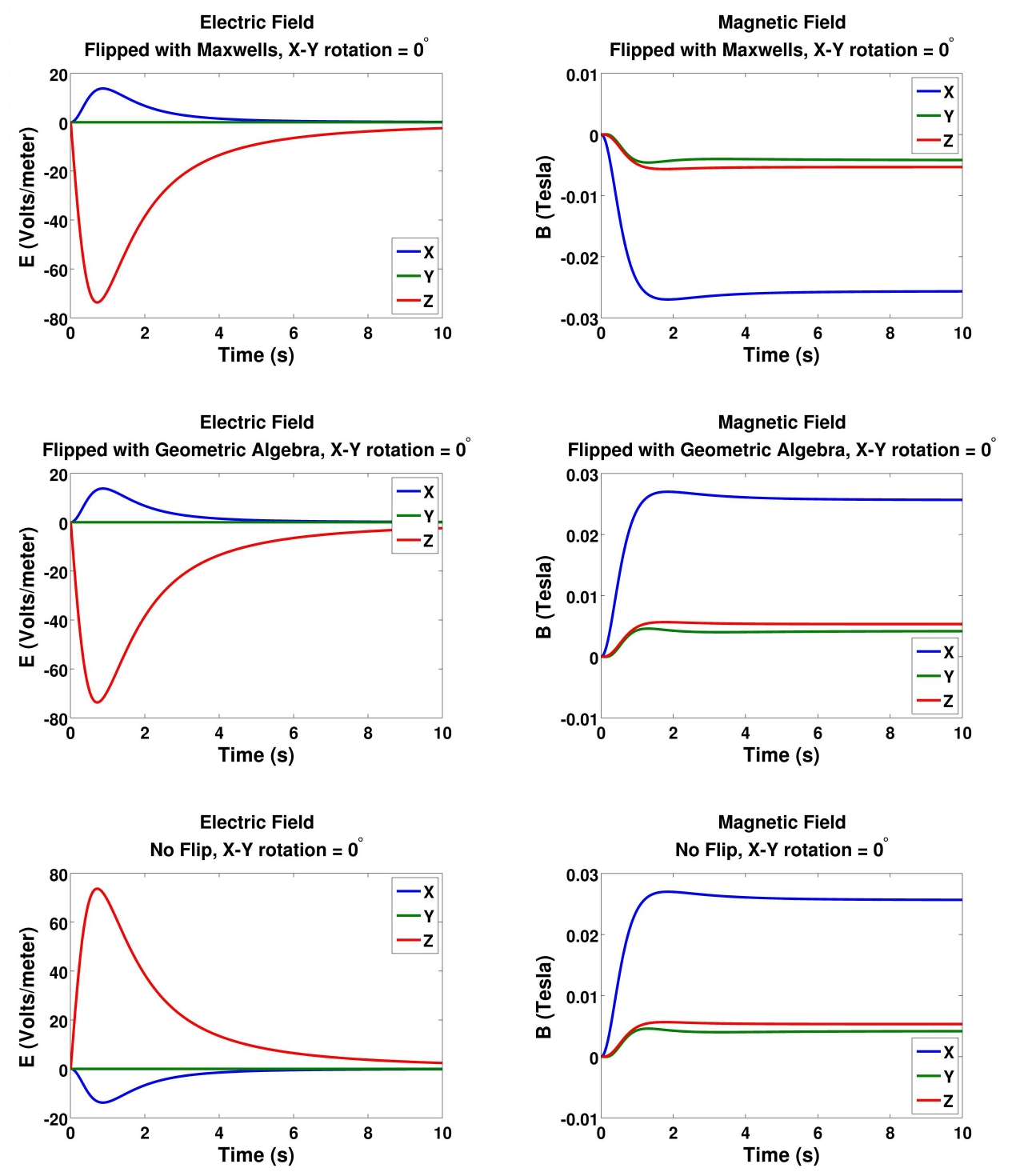}\\
    \label{fig:fig03}
    \caption{Initial case of the point-charge study, with the inversion in all three directions, where ${e_1}$ = -${e_1}$, ${e_2}$ = -${e_2}$, and ${e_3}$ = -${e_3}$.}
\end{figure}

\begin{figure}
\centering
\includegraphics[width=\textwidth]{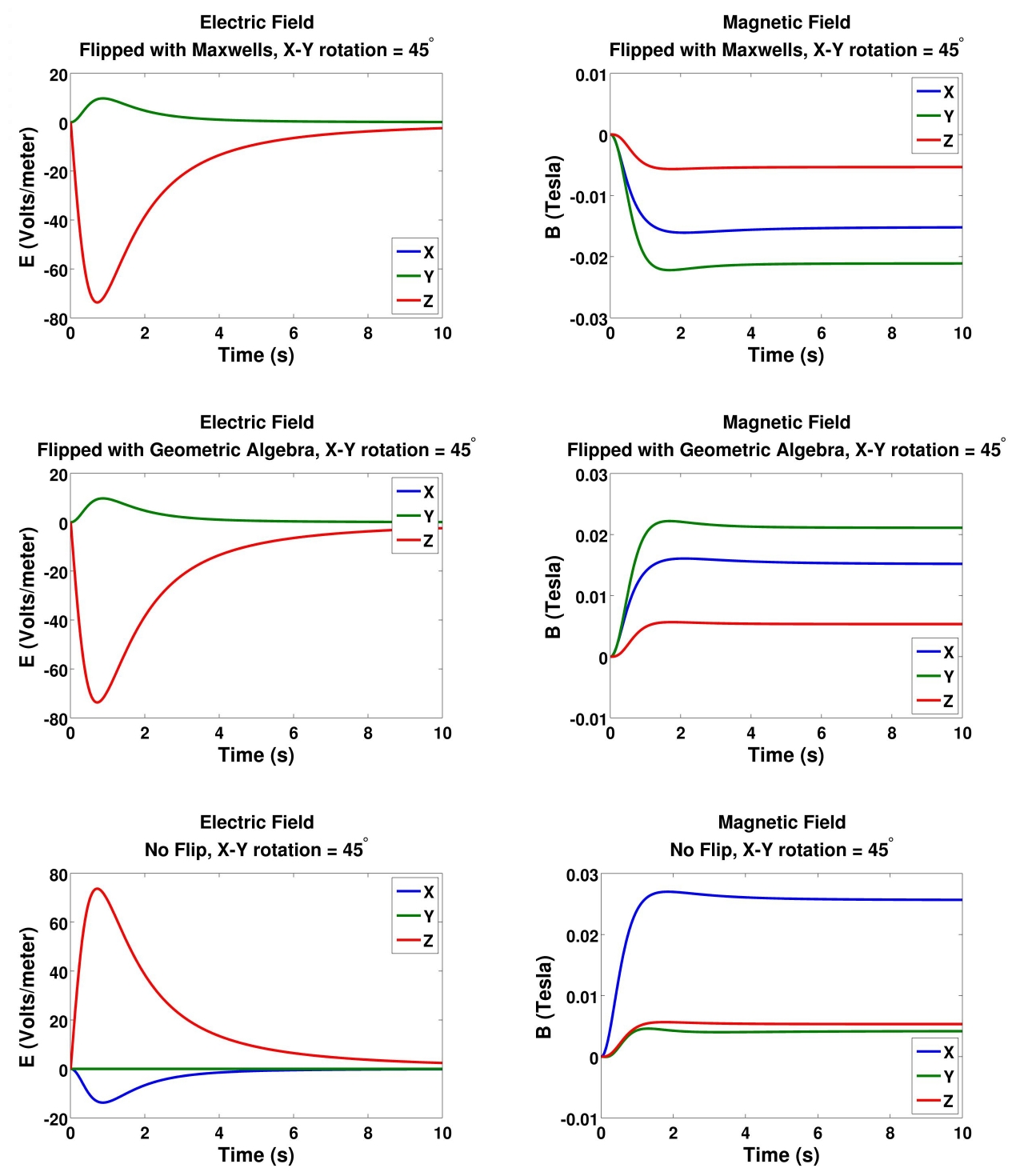}\\
    \label{fig:fig04}
    \caption{The second case, with the inversion in all three directions, where ${e_1}$ = -${e_1}$, ${e_2}$ = -${e_2}$, and ${e_3}$ = -${e_3}$, as well as the rotation along the X-Y plane of $\phi$ = 45$^{\circ}$.  Note, with the flipped electric field, ${{E_X}={E_Y}}$.}
\end{figure}

So far, the electric and magnetic fields were determined with Maxwell's equations, at a fixed reference point.  The first study (Figure \ref{fig:fig03}) was simply inverted (${e_1}$ = -${e_1}$, ${e_2}$ = -${e_2}$, and ${e_3}$ = -${e_3}$).  The second study (Figure \ref{fig:fig04}) will have the above inversion, ${e_1}$ = -${e_1}$, ${e_2}$ = -${e_2}$, and ${e_3}$ = -${e_3}$, and then rotate the reference point by $\phi$ = 45$^{\circ}$.  \\

R=
$\begin{bmatrix}
{-1} & {0} & {0}\\
{0} & {-1} & {0}\\
{0} & {0} & {-1}\\
\end{bmatrix}
{\times}
\begin{bmatrix}
{cos(\phi)} & {-sin(\phi)} & {0}\\
{sin(\phi)} & {cos(\phi)} & {0}\\
{0} & {0} & {1}\\
\end{bmatrix}
=
\begin{bmatrix}
{-{\sqrt{2}}/2} & {{\sqrt{2}}/2} & {0}\\
{-{\sqrt{2}}/2} & {-{\sqrt{2}}/2} & {0}\\
{0} & {0} & {-1}\\
\end{bmatrix}$
.

\section{Moving Observer Study \#1}

\begin{figure}
\centering
\includegraphics[width=\textwidth]{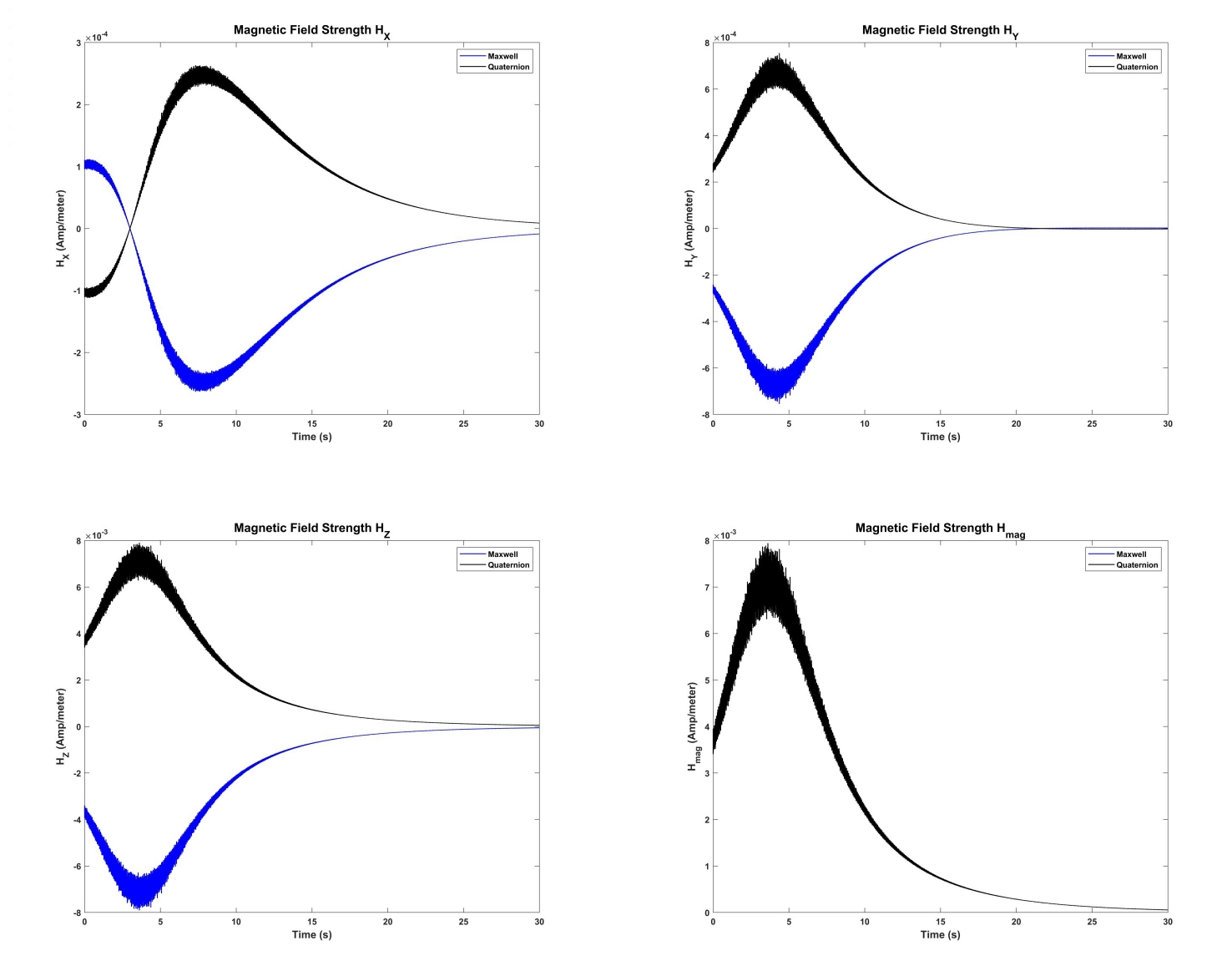}\\
\label{fig:fig05}
\caption{Results of moving-observer study.}
\end{figure}

A parametric study was done, to determine if one would observe a different magnetic field between two moving observers, and if quaternion geometry could detect these anomalies.  In the moving observer study (Figure \ref{fig:fig05}), the observer has an X-Y-Z acceleration of (-10, 10, 1) m/s${^2}$, a velocity of (-2, 200, 0) m/s, and an initial position of (-1000, +1000, 300) m.  The target has no acceleration, has a velocity of (1, 1, 0.1) m/s, and is initially located at (0, 0, -20) meters.  The goal was to roughly suggest an aircraft monitoring a submarine.  The target is treated as a solenoid with a steel core ($\mu_R$ = 2,000), 10 meters in radius, 100 meters in length, with an equivalent current of 1 amp flowing through the solenoid (hull).  

Fractal geometry is a component of this study, as a form of adding randomness typically seen in nature, to confirm if the quaternion geometry can properly capture it.  A fractal geometry profile used typically for surface roughness and friction \cite{Stachowiak_EngTrib} was used, where the variation $\alpha$ was determined
\begin{eqnarray}
{{\alpha}(x)}&=&{{(Ra^{D-1})}{\cdot}{{\Sigma_j}{\frac{cos(2{\cdot}{\pi}{\cdot}{\gamma^j}{\cdot}x)}{{{\gamma}^{(2-D){\cdot}j}}}}}},\nonumber
\end{eqnarray}
where \emph{D} = 1.6, $\gamma$ = 1.5, and \emph{Ra} = 108.221; these values are arbitrary based on the roughness profile desired.  The position in the \emph{x}, \emph{y}, and \emph{z} direction between the observer and the target was consistently fluctuated by 50${\cdot}{\alpha}$ meters, with \emph{x} ranging from 1 to 100,000 before repeating itself, and \emph{j} ranging from 1 to 100.  

Finally, one important step is to determine the rotation matrix for the changing reference plane.  This is determined by comparing the velocity profiles of the observer (${V_{xo}}$, ${V_{yo}}$, ${V_{zo}}$) and the target (${V_{xt}}$, ${V_{yt}}$, ${V_{zt}}$).  The first step is to calculate the dot product between the observer and the target; if it is negative, the target is inverted for the time being.  The following formula is used to find the rotation matrix

\begin{eqnarray}
X_{12}&=&\sqrt{{\Sigma}{({V_o}{\times}{V_t})^2}},\nonumber
\end{eqnarray}\\

G=
$\begin{bmatrix}
{{V_o}{\bullet}{V_t}} & {X_{12}} & {0}\\
{X_{12}} & {{V_o}{\bullet}{V_t}} & {0}\\
{0} & {0} & {1}\\
\end{bmatrix}$

\begin{eqnarray}
u&=&V_o,\nonumber \\
v&=&\frac{{V_t}-{({V_o}{\bullet}{V_t}){\cdot}{V_o}}}{\sqrt{{\Sigma}{V_t}-{(({V_o}{\cdot}{V_t}){\cdot}{V_o})^2}}},\nonumber \\
w&=&{V_o}{\times}{V_t},\nonumber \\
F&=&\frac{1}{u{\cdot}v{\cdot}w},\nonumber \\
{\bf{RR}}&=&{F^{-1}}{\cdot}{({G}{\cdot}{F})},\nonumber
\end{eqnarray}
If the initial dot product was negative, and the target is inverted, then the rotation matrix is made negative as well\\

${\bf{RR}}={F^{-1}}{\cdot}{({G}{\cdot}{F})}=
\begin{bmatrix}
{-1} & {0} & {0}\\
{0} & {-1} & {0}\\
{0} & {0} & {-1}\\
\end{bmatrix}$
. 

This simulation is run for 30 seconds with a time-step of a millisecond (30,000 time-steps).  The results of the magnetic field H (amp/meter) are plotted in Figure 7, both for the Maxwell and Quaternion approach.  It was observed, with the varying direction, the magnetic fields in the X (Figure 7a), Y (Figure 7b), and Z (Figure 7c) direction are inverted; this phenomenon is only observed when the velocities are inverted.  The magnitude of both approaches (Figure 7d) matches perfectly, as expected.  This simulation shows how quaternion geometry can accurately capture the polar vector properties of magnetic fields.  

\section{Moving Observer Study \#2}

\subsection{Without fractal geometry}

Another parametric study was developed, with an observer (ex an aircraft) and a target (ex a submarine).  The observer is flying back and forth in a pattern as it is surveilling the magnetic fields from the target, which is moving in a circular direction.  The observer aircraft starts off at a position of \emph{X} = 100,000 meters, \emph{Y} = -100,000 meters, and a height of 1,000 meters.  The target submarine starts at \emph{X} = 100,000 meters, \emph{Y} = -50,000 meters, and a depth of -20 meters.  The target submarine treated as magnetic field in a solenoid, with a current \emph{I} = 1, and a relative magnetic permeability of $\mu_R$ = 2000.  The observer travels in a straight path for 300 seconds (with a time-step of 0.1 second), 
\begin{eqnarray}
{V_{X,O}'}&=&{D{\cdot}{100}},\nonumber
\end{eqnarray}
where \emph{D} is either 1 or -1, and then it turns over 300 seconds, at a velocity of: 
\begin{eqnarray}
{V_{X,O}'}&=&{D{\cdot}{100}{\cdot}{cos({\frac{2{\cdot}{\pi}{\cdot}{t'}}{300}})}},\nonumber \\ \nonumber
{V_{Y,O}'}&=&{{100}{\cdot}{sin({\frac{2{\cdot}{\pi}{\cdot}{t'}}{300}})}}.
\end{eqnarray}
where \emph{t'} (s) resets every 600 seconds.  The Z-position of the observer is also sinusoidal, following the velocity profile of 
\begin{eqnarray}
{V_{Z,O}'}&=&{{0.2}{\cdot}{cos({\frac{2{\cdot}{\pi}{\cdot}{t}}{1200}})}}.\nonumber
\end{eqnarray}
The target submarine, on the other hand, follows the following function for ${V_{X,T}'}$ (m/s), ${V_{Y,T}'}$ (m/s), and ${V_{Z,T}'}$ (m/s)
\begin{eqnarray}
{V_{X,T}'}&=&{{40}{\cdot}{cos({\frac{2{\cdot}{\pi}{\cdot}{t}}{9000}})}},\nonumber \\ \nonumber
{V_{Y,T}'}&=&{{40}{\cdot}{sin({\frac{2{\cdot}{\pi}{\cdot}{t}}{9000}})}},\nonumber \\ \nonumber
{V_{Z,T}'}&=&{{0.1}{\cdot}{cos({\frac{2{\cdot}{\pi}{\cdot}{t}}{900}})}}.
\end{eqnarray}

\begin{figure}
\centering
\includegraphics[width=\textwidth]{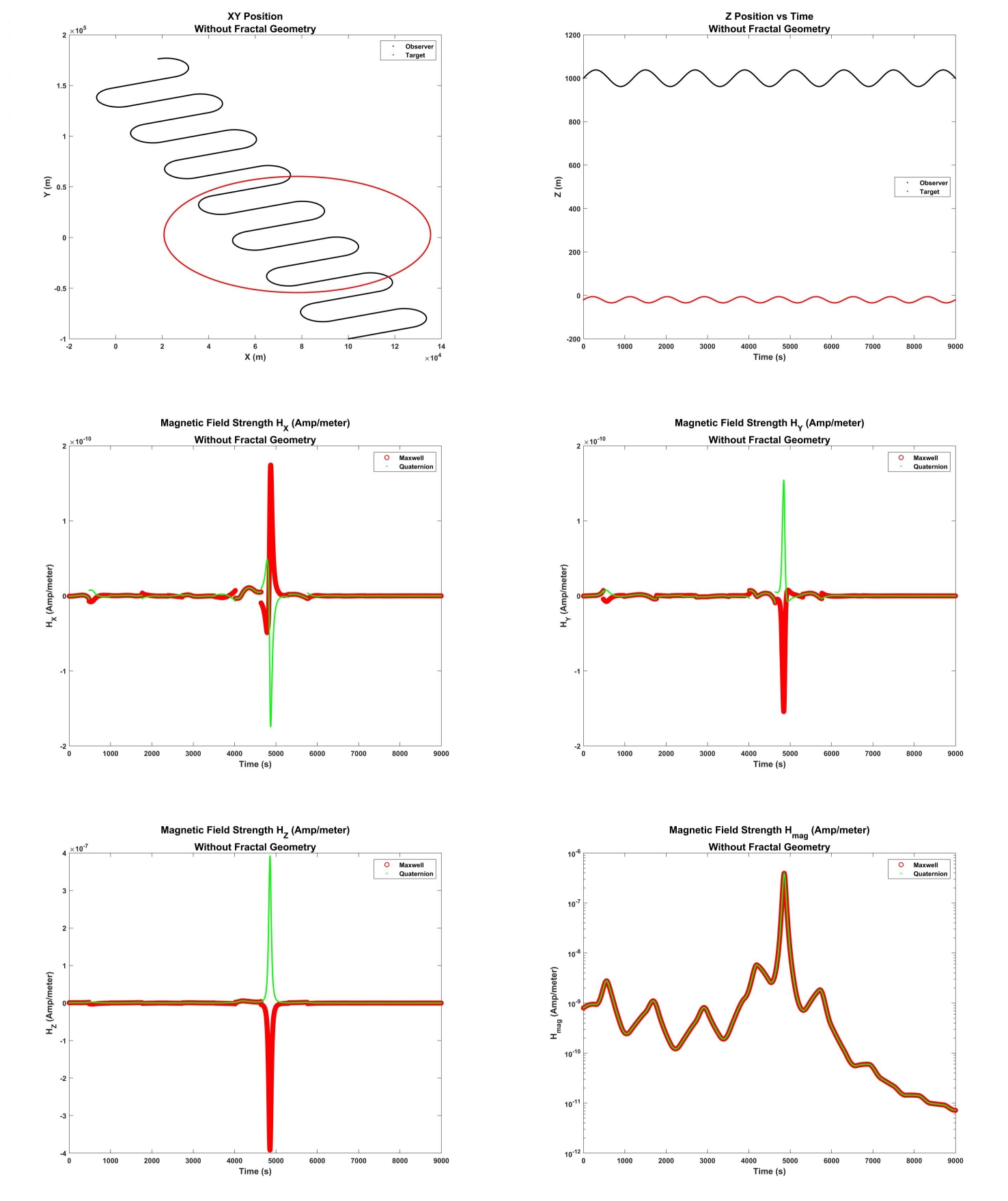}\\
    \label{fig:fig06}
    \caption{(a) \emph{X}, \emph{Y}, and (b) \emph{Z} position of target and observer.  Magnetic field strength from target as detected by observer, in the (c) \emph{X}, (d) \emph{Y}, (e) \emph{Z} direction, as well as (f) the magnitude, without fractal geometry.}
\end{figure}

The velocities of both the target and the observer is rotated along the \emph{Z}-axis by 22.5$^{\circ}$.  The \emph{X} and \emph{Y} position of the observer and target are plotted in Figure \ref{fig:fig06}-a, and the \emph{Z} position in Figure \ref{fig:fig06}-b.  The magnetic field strength {\bf{H}} (Amp/meter) is plotted in the \emph{X} (Figure \ref{fig:fig06}-c), \emph{Y} (Figure \ref{fig:fig06}-d), and \emph{Z} (Figure \ref{fig:fig06}-e) direction, as well as the magnetic field strength magnitude (Figure \ref{fig:fig06}-f).  It is clear that using quaternion geometry will provide a different direction for the magnetic field strength, but at the correct \emph{X-Y-Z} position; using traditional Maxwell's will give an erroneous direction for the magnetic field strength.  The magnitude of the magnetic field H is consistent with both Maxwell's and Quaternion Geometry.  

\subsection{With fractal geometry}

\begin{figure}
\centering
\includegraphics[width=\textwidth]{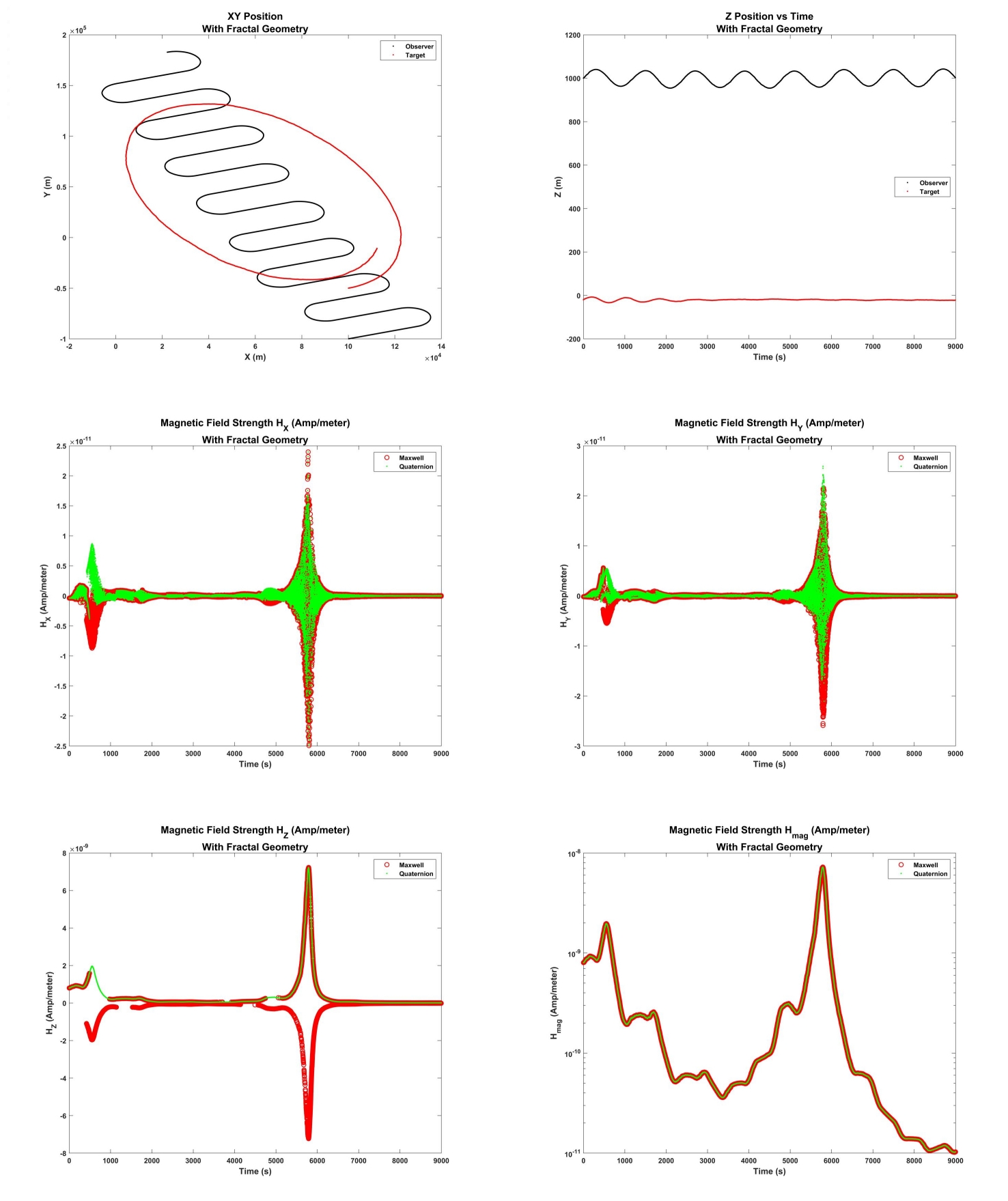}\\
    \label{fig:fig07}
    \caption{(a) \emph{X}, \emph{Y}, and (b) \emph{Z} position of target and observer.  Magnetic field strength from target as detected by observer, in the (c) \emph{X}, (d) \emph{Y}, (e) \emph{Z} direction, as well as (f) the magnitude, with fractal geometry.}
\end{figure}

This simulation was repeated, but with fractal geometry altering the velocity.  
\begin{eqnarray}
{{\alpha}(x,y)}&=&1+{{{0.01}{\cdot}{[{(Ra^{D-1})}{\cdot}{{\Sigma_{j=1}^10}{\frac{{cos(2{\cdot}{\pi}{\cdot}{\gamma^j}{x})}{\cdot}{cos(2{\cdot}{\pi}{\cdot}{\gamma^j}{x})}}{\gamma^{(2-D){\cdot}{j}}}]}}}}},\nonumber
\end{eqnarray}
where \emph{D} = 1.6, $\gamma$ = 1.5, and \emph{Ra} = 80.  When both the observer and the target were moved in position, the value of $\alpha$ was altered to the velocity, where $X={X_0}+{V{\cdot}{\alpha}{\cdot}{\delta}t}$.  In addition, the target used fractal geometry to generate an imperfect circular motion, where ${\theta}={|\alpha|}^{1.1}/{\alpha}$.  
\begin{eqnarray}
{V_{X,T}'}&=&{40}{\cdot}{cos(\frac{2{\cdot}{\pi}{\cdot}{\theta}{\cdot}t}{9000})},\nonumber \\
{V_{Y,T}'}&=&{40}{\cdot}{sin(\frac{2{\cdot}{\pi}{\cdot}{\theta}{\cdot}t}{9000})},\nonumber \\
{V_{Z,T}'}&=&{0.1}{\cdot}{cos(\frac{2{\cdot}{\pi}{\cdot}{\theta}{\cdot}t}{900})}.\nonumber
\end{eqnarray}
The positions and magnetic field strength for this simulation with fractal geometry are all plotted in Figure \ref{fig:fig07}.

\section{Eddy Current Brake Model}

An eddy current brake \cite{Eddy01_01, Eddy03_01_good, Eddy01_02, Eddy02_01, Eddy02_02, Eddy02_03, Eddy02_04, Eddy03_02, Eddy03_03, Eddy03_04, Eddy04_01, Eddy04_02} involves a moving, conductive, non-ferrous material (ex. copper, aluminum, gold) moving within a magnetic field.  A finite difference model was built to simulate this phenomenon.  The first step is to calculate the magnetic field strength \emph{H} (Amp / meter), using the Biot-Savart law (equation \ref{eq:eqBSlaw})
\begin{eqnarray}
\label{eq:eqBSlaw}
{\bf{H}}&=&{\frac{I}{4{\cdot}{\pi}{R^3}}}{\cdot}{{\int}{{d{\bf{I}}}{\times}{\bf{R}}}}.
\end{eqnarray}
If looking for the magnetic field strength of a single coil, the change in length of the current ${d{\bf{I}}}$ is ${\rho}{\cdot}{d{\theta}}$, and the equation for the magnetic field strength at a given position away is thus 
\begin{eqnarray}
{\bf{H}}&=&{\frac{I{\cdot}{\rho^2}}{2{\cdot}{({R^2}+{\rho^2})^{\frac{3}{2}}}}}{\bf{a_z}},
\end{eqnarray}
where $\rho$ (m) represents the radius of the coil, and \emph{R} (m) represents the distance from the point of {\bf{H}} to the center of the coil.  In this model, the user will specify the number of coils, and the number of coils per unit distance, and the model will evenly space out incremental circular loops, and combine the total magnetic field strength at every distance increment.  After this has been calculated, the model will determine the magnetic field {\bf{B}} by multiplying the magnetic permeability $\mu$.  So long as the disk is a non-ferrous conductive metal (ex copper, aluminum), it is reasonable to treat the magnetic permeability as the magnetic permeability of a vacuum ${\mu_0}=4{\cdot}{\pi}{\cdot}{10^{-7}}$.  

When a conductor moves through a magnetic field, electric eddy currents are induced, and a Lorentz force is applied.  The induced current density {\bf{J}} (Amps/m${^3}$) is 
\begin{eqnarray}
{\bf{J}}&=&{{\sigma}{\cdot}{({\bf{b}}{\times}{\bf{B}})}},
\end{eqnarray}
where {\bf{v}} (m/s) is the velocity vector.  The model is a rotating disk on the \emph{X}-\emph{Y} plane, therefore the velocity at a given point is thus
\begin{eqnarray}
{V_X}&=&{{\Omega}_{RPM}}{\cdot}{\frac{2{\cdot}{\pi}}{60}}{\cdot}{R_{XY}}{\cdot}{|cos(\theta)|}{\cdot}{\frac{xx}{|xx|}},\\ \nonumber
{V_Y}&=&{{\Omega}_{RPM}}{\cdot}{\frac{2{\cdot}{\pi}}{60}}{\cdot}{R_{XY}}{\cdot}{|sin(\theta)|}{\cdot}{\frac{yy}{|yy|}},\\ \nonumber
{\theta}&=&{tan^{-1}(\frac{yy}{xx})}.
\end{eqnarray}

The Lorentz force is simply applied as a power consumption \emph{P} (Watts) as
\begin{eqnarray}
P&=&{{\frac{J^2}{\sigma}}{\cdot}{V}},\\ \nonumber
{J^2}&=&{J_X^2}+{J_Y^2}+{J_Z^2}={J_X^2}+{J_Y^2}, 
\end{eqnarray}
where \emph{V} is the volume (m${^3}$) of a given grid node.  Once the power for all of the grid nodes is calculated and summed up, the torque is calculated simply by dividing the angular torque $\omega$ (rad/s)
\begin{eqnarray}
T&=&{\frac{P}{\omega}},\\ \nonumber
{\omega}&=&{\Omega_{RPM}}{\cdot}{\frac{2{\cdot}{\pi}}{60}}.
\end{eqnarray}

\subsection{Limiting Speed}

There is a limit on the effectiveness of the eddy current brake with increasing rotor disk speed.  As the disk moves through the magnetic field, eddy currents are generated proportional to both the disk speed and the conductivity of the disk material; these currents induce a magnetic field of their own that counteracts the magnetic field of the solenoid.  Previous references  have cited the limited linear speed $V_{lim}$ (m/s) before an eddy current brake loses its effectiveness as 
\begin{eqnarray}
{V_{lim}}&=&{\frac{2}{{d_{disk}}{\cdot}{{\sigma}_{disk}}{\cdot}{{\mu}_{disk}}}},
\end{eqnarray}
where $d_{disk}$ (m) is the thickness of the disk, ${\sigma}_{disk}$ (1/ohms) is the electrical conductivity of the disk material, and ${\mu}_{disk}$ (H/m) is the magnetic permeability of the disk material.  It is based on the limit where the magnetic field induced by the eddy currents reach twice the magnet field of the solenoid, or 
\begin{eqnarray}
{\frac{{\bf{B}}_{eddy}}{{\bf{B}}_{solenoid}}}&=&2.
\end{eqnarray}
As demonstrated in the analytical derivation, the eddy currents $I_{eddy}$ (amps) is
\begin{eqnarray}
{I_{eddy}}&=&{{{\delta}t}_{disk}}{\cdot}{{\sigma}_{disk}}{\cdot}{|{\bf{v}}{\times}{\bf{B}}|},
\end{eqnarray}
and the magnetic field induced by the eddy current is roughly ${\bf{B}}_{eddy}{\approx}{\mu_0}{\cdot}{I_{eddy}}$, and therefore a relationship between ${\bf{B}}_{eddy}$ and ${\bf{B}}_{solenoid}$ can be found as 
\begin{eqnarray}
{B_{eddy}}&{\approx}&{{{\mu}_0}{\cdot}{{\delta}t_{disk}}{\cdot}{{\sigma}_{disk}}{\cdot}{V}{\cdot}{B_{solenoid}}}.
\end{eqnarray}
If one sets the limit of ${B_{eddy}}/{B_{solenoid}}=2$ for the limiting speed, then the above relationship can be rewritten as
\begin{eqnarray}
{B_{eddy}}/{B_{solenoid}}=2&=&{{\mu}_0}{\cdot}{{\delta}t_{disk}}{\cdot}{{\sigma}_{disk}}{\cdot}{V_{LIM}},\\ \nonumber
{V_{LIM}}&=&{\frac{2}{{{\mu}_0}{\cdot}{{\delta}t_{disk}}{\cdot}{{\sigma}_{disk}}}}.
\end{eqnarray}
This limit is purely theoretical, and many factors, including the electromagnetic solenoid design and placement, will ultimately affect the magnetic field.  Single examples of a countering eddy current magnetic field exceeding twice the solenoid magnetic field does not necessarily mean a total reduction in torque.  For this reason, a numerical model is the optimal way to investigate this balance.

\subsection{Basic Eddy Current Model}

This model assumes an electromagnet, defined as a solenoid on both sides of a disk.  

\begin{figure}
\centering
\includegraphics[width=\textwidth]{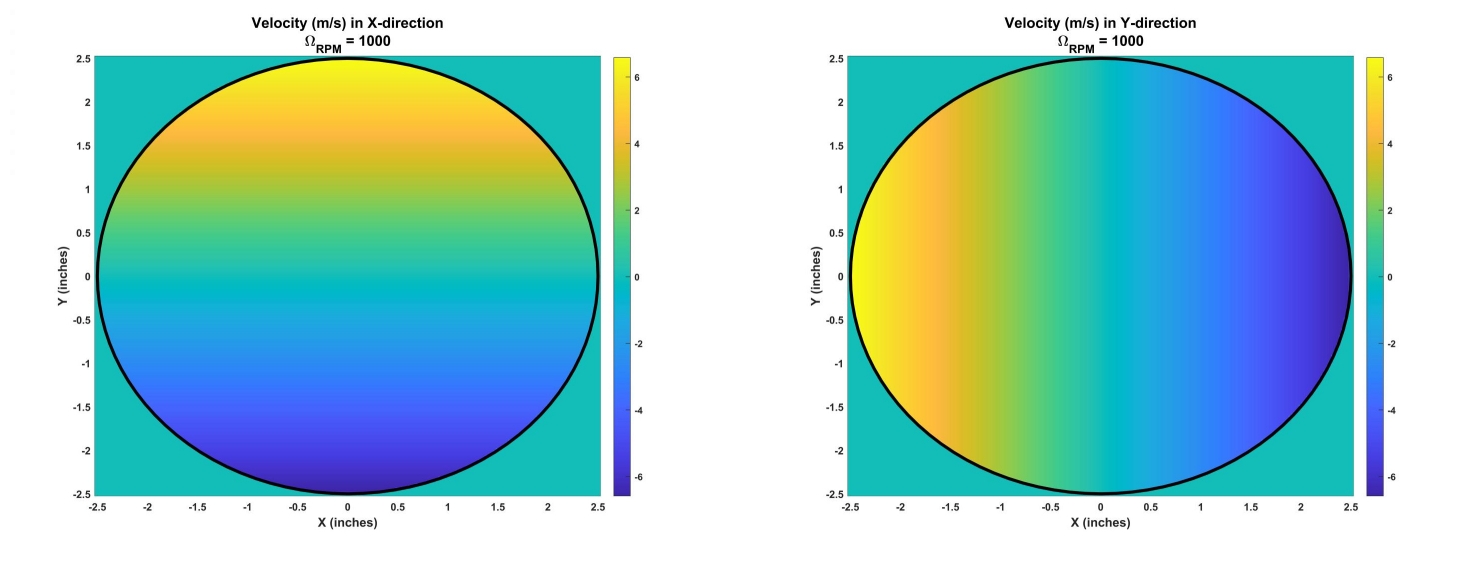}\\
    \label{fig:fig08}
    \caption{Velocity (m/s) profile of spinning 5” diameter copper disk at 1,000 RPM, in the (a) X and (b) Y direction.}
\end{figure}

\begin{figure}
\centering
\includegraphics[width=\textwidth]{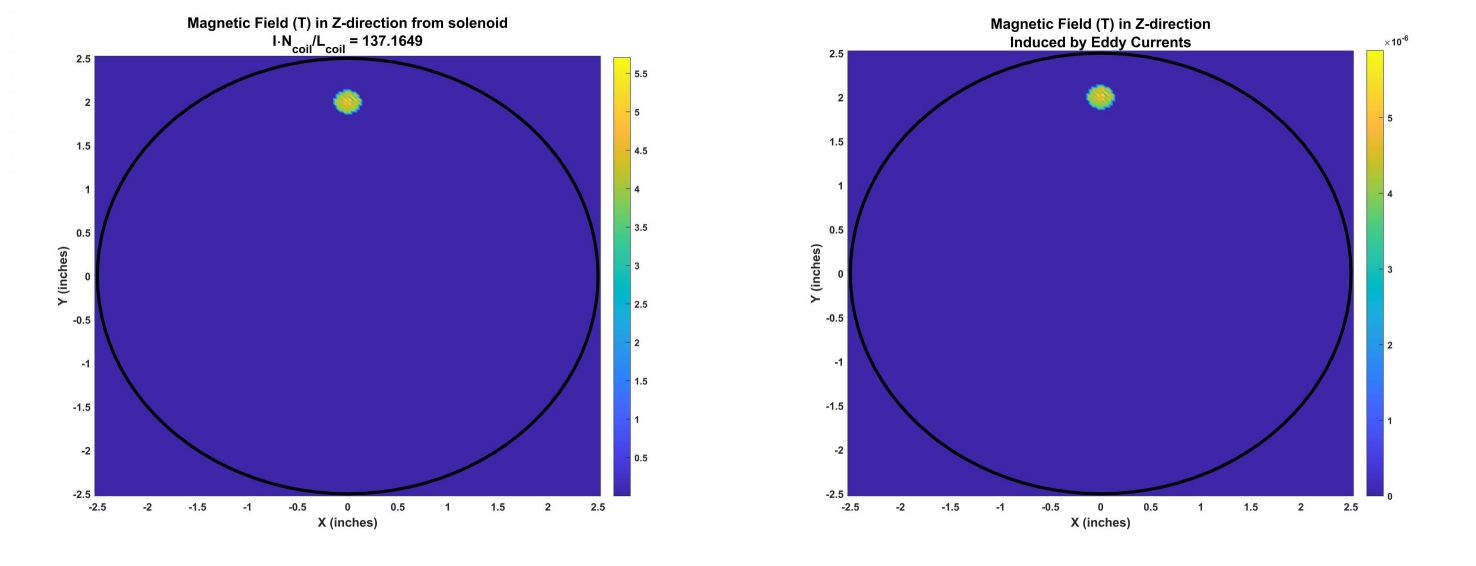}\\
    \label{fig:fig09}
    \caption{Comparison of {\bf{B}}, both (a) solely from the solenoid, versus (b) magnetic field generated from the eddy currents.}
\end{figure}

\begin{figure}
\centering
\includegraphics[width=\textwidth]{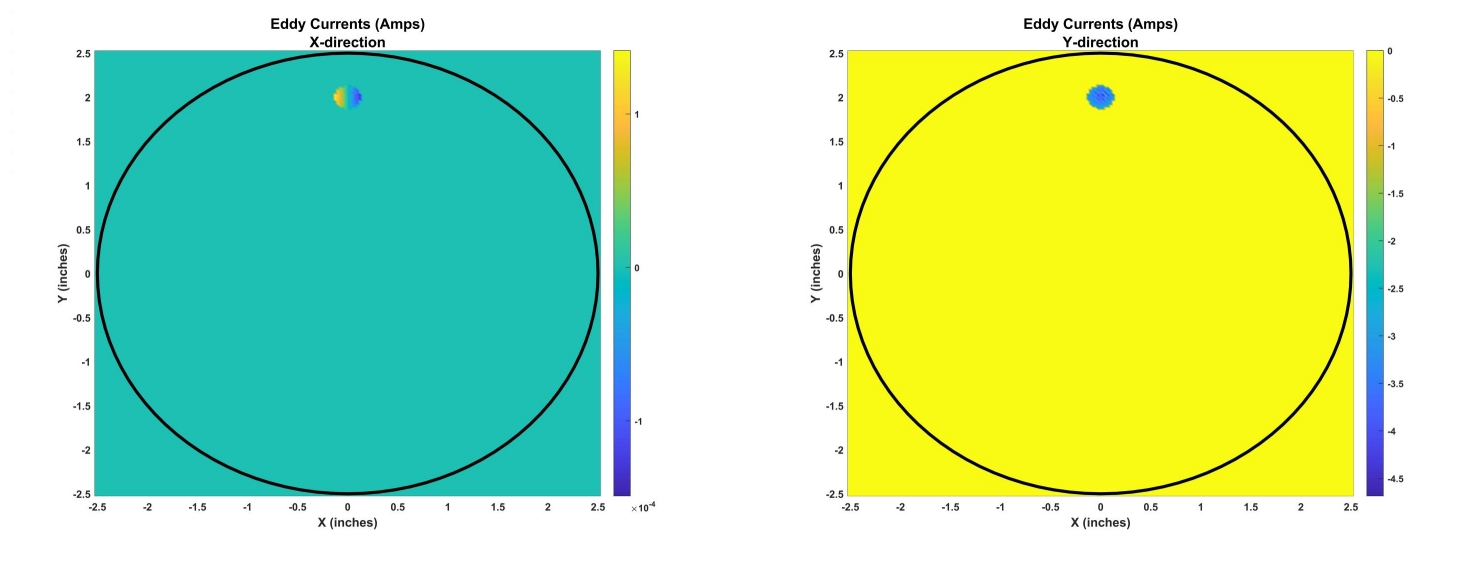}\\
    \label{fig:fig10}
    \caption{Eddy Current generated (amps), in the (a) X and (b) Y direction.}
\end{figure}

\begin{figure}
\centering
\includegraphics[width=\textwidth]{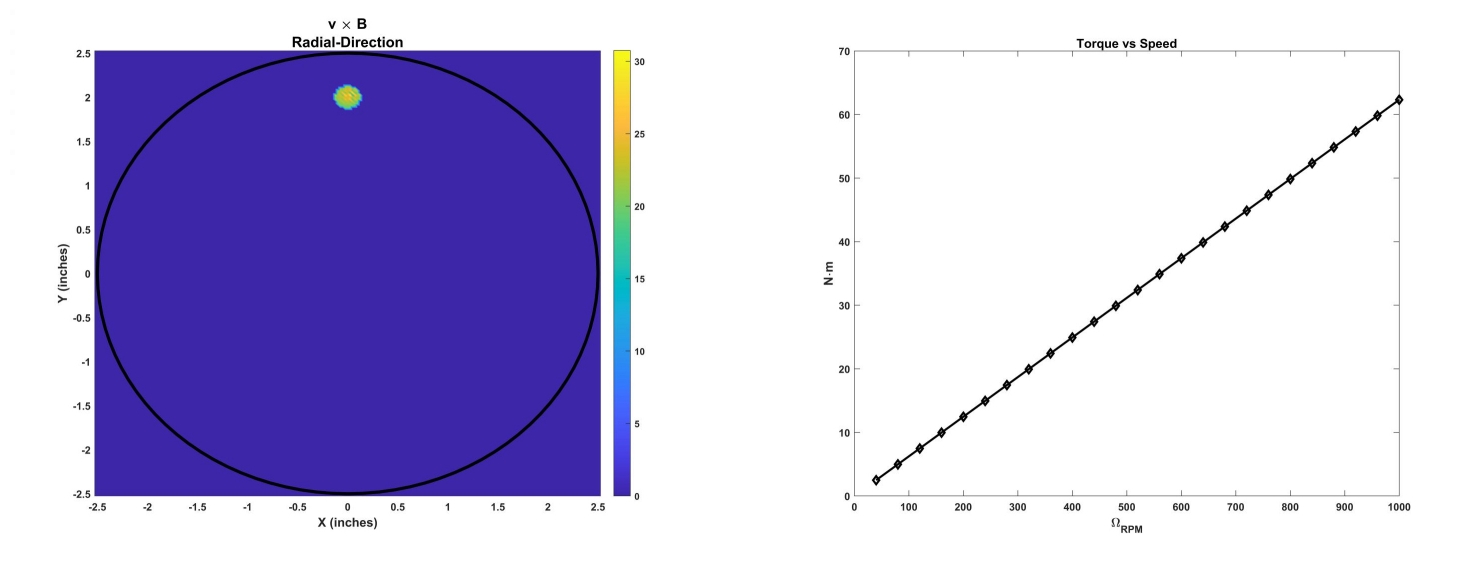}\\
    \label{fig:fig11}
    \caption{(a) the cross product between the radial velocity vector {\bf{v}} (m/s) of the rotating disk (1,000 RPM) and the magnetic field {\bf{B}} (T), and (b) the torque as a function of disk speed (RPM).  }
\end{figure}

The simulation depicted in Figure \ref{fig:fig08}, Figure \ref{fig:fig09}, Figure \ref{fig:fig10}, and Figure \ref{fig:fig11} tests an eddy current brake, utilizing a 5 inch diameter, 0.25 inch thick copper disk, rotating at 1,000 RPM.  One solenoid coil of 22 loops each, 0.260 inch in diameter, over 3.15 inch of length with a 99.95\% iron annealed in hydrogen (${\mu}_R$ = 200,000, ${\epsilon}_R$ = 2) core, is situated 0.1525 inch above and below the copper disk, at a radius of 4.5 inch on the X-axis position.  A current of 0.7 amps is flowing through the solenoid coil in this simulation.  

As the velocity increases, the eddy currents increase in magnitude, and these currents (Figure \ref{fig:fig10}) result in a countering magnetic field (Figure \ref{fig:fig09}-b).  These countering magnetic fields reduce the net magnetic field, and thus when as the rotating speed of the disk increases, this countering effect gets greater and greater, until eventually there is a limit to the torque generated with greater speed.  

The model demonstrated that for a constant speed, the torque is approximately proportional to the solenoid current (and the ratio of the solenoid coils per length) squared; for a constant current, the torque is approximately linearly proportional to the disk speed (until the eddy-current induced magnetic field becomes significant).  An empirical equation (correlation coefficient \emph{R} = 1.00) for the numerically obtain torque \emph{T} (N$\cdot$m) with this designed eddy current break is
\begin{eqnarray}
T&=&{3.3125}{\cdot}{10^{-6}}{\cdot}{{({I}{\cdot}{\frac{N_{coil}}{L_{coil}}})}^2}{\cdot}{{\Omega}_{RPM}}.
\end{eqnarray}
This equation is applicable up till 1,000 RPM; at substantially higher speeds (not realistic with the applications of interest) the torque increase will decrease.

\subsection{Eddy Current Probe Study}

Another parametric study was conducted with the eddy current electric brake, where the 40" diameter, 1/2" thick copper disk is spinning at 1,000 RPM.  There are two 4 inch radius magnetic solenoids, 180$^{\circ}$ apart, situated 1/16 inch above and below the disk.  The parametric study is analyzing the magnetic and electric fields measured in the disk, as situated 1/16 inch \emph{above} the spinning disk, parametrically moving in the \emph{X} and \emph{Y} direction (53$\times$53 grid points).  The parametric study is performed three times, where the reference point is directly in the \emph{Z} direction (0, 0, 1), where the reference point rotates with position (0, 0, 1), where the position rotates with position (\emph{X}, \emph{Y}, 1), and when there is a random wobble ($\pm$0.05, $\pm$0.05, 1$\pm$0.05).  

\begin{figure}
\centering
    \includegraphics[width=\textwidth]{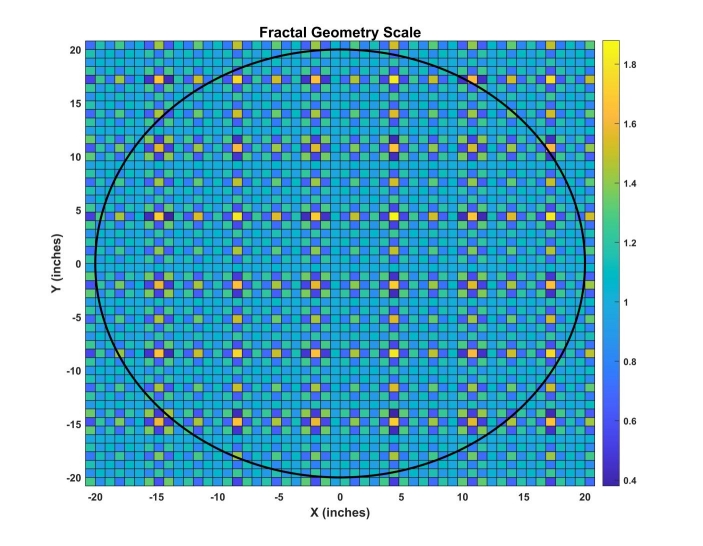}
    \label{fig:fig12}
    \caption{Weighting function for Fractal Geometry.}
\end{figure}

The parametric studies are run both without and with fractal geometry \cite{Stachowiak_EngTrib, Turbulence_01, Turbulence_02}, to simulate random currents within the disk.  This is conducted by apply a weighting factor over the 53x53 grid, using the following function: 
\begin{eqnarray}
{S(x,y)}&=&{[{{\Sigma}_{a=1}^N}{cos({\frac{2{\cdot}{\pi}{\cdot}{{\gamma}^a}{\cdot}{x}}{{\gamma}^{(2-D){\cdot}{a}}})}}]}{\cdot}{[{{\Sigma}_{b=1}^N}{cos({\frac{2{\cdot}{\pi}{\cdot}{{\gamma}^b}{\cdot}{y}}{{\gamma}^{(2-D){\cdot}{b}}})}}]}{\cdot}{Ra^{2{\cdot}{(D-1)}}},
\end{eqnarray}
where \emph{N} = 1,000, $\gamma$ = 1.5, \emph{D} = 0.6, and \emph{Ra} = 1,000.  This function was then scaled, so as to have a mean of 1, and a standard deviation of 10.  The results are plotted in Figure \ref{fig:fig12}.  

\begin{figure}
\centering
\begin{minipage}[b]{0.85\linewidth}
\centering
    \includegraphics[width=\textwidth]{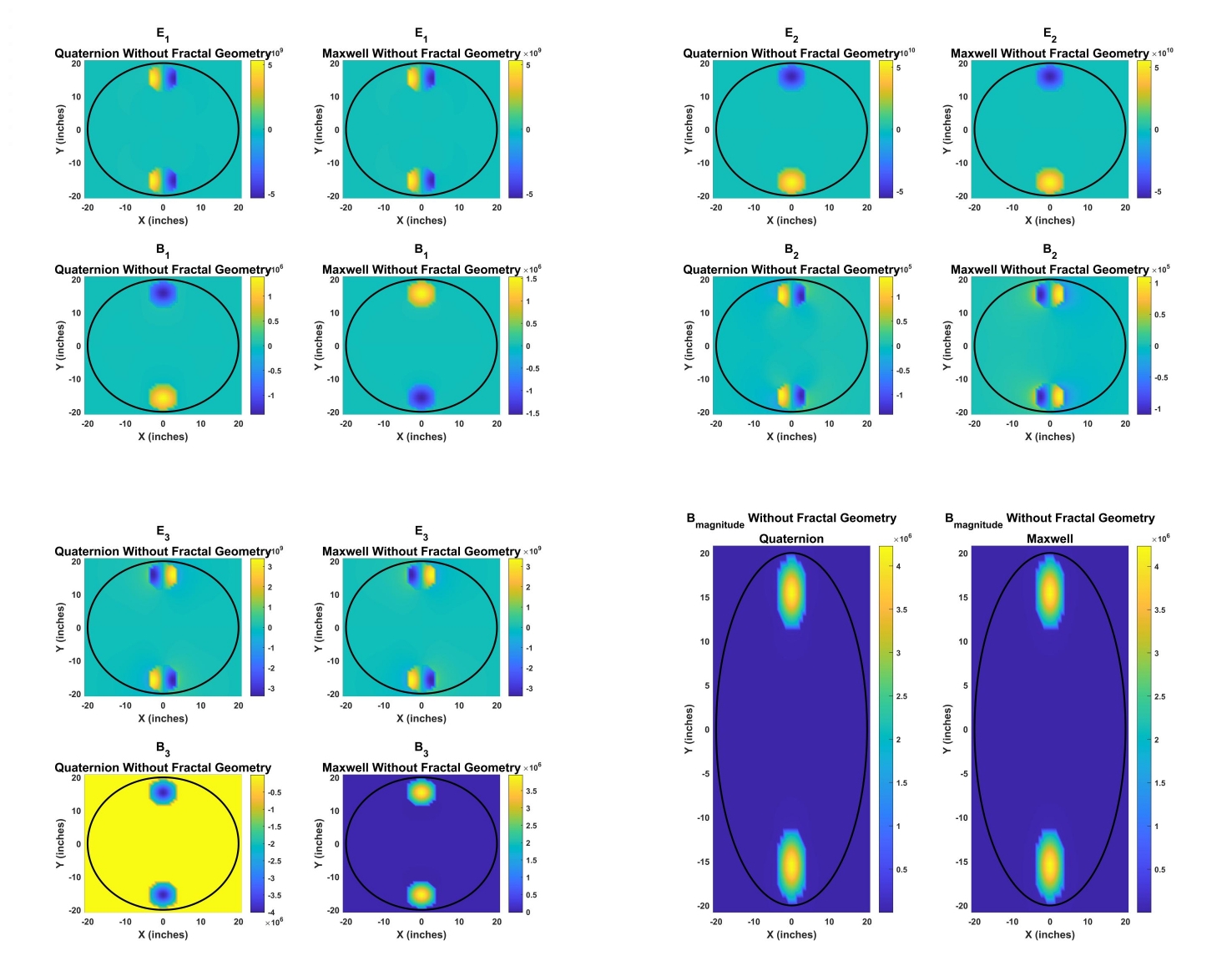}\\
\end{minipage}
\begin{minipage}[b]{0.85\linewidth}
\centering
    \includegraphics[width=\textwidth]{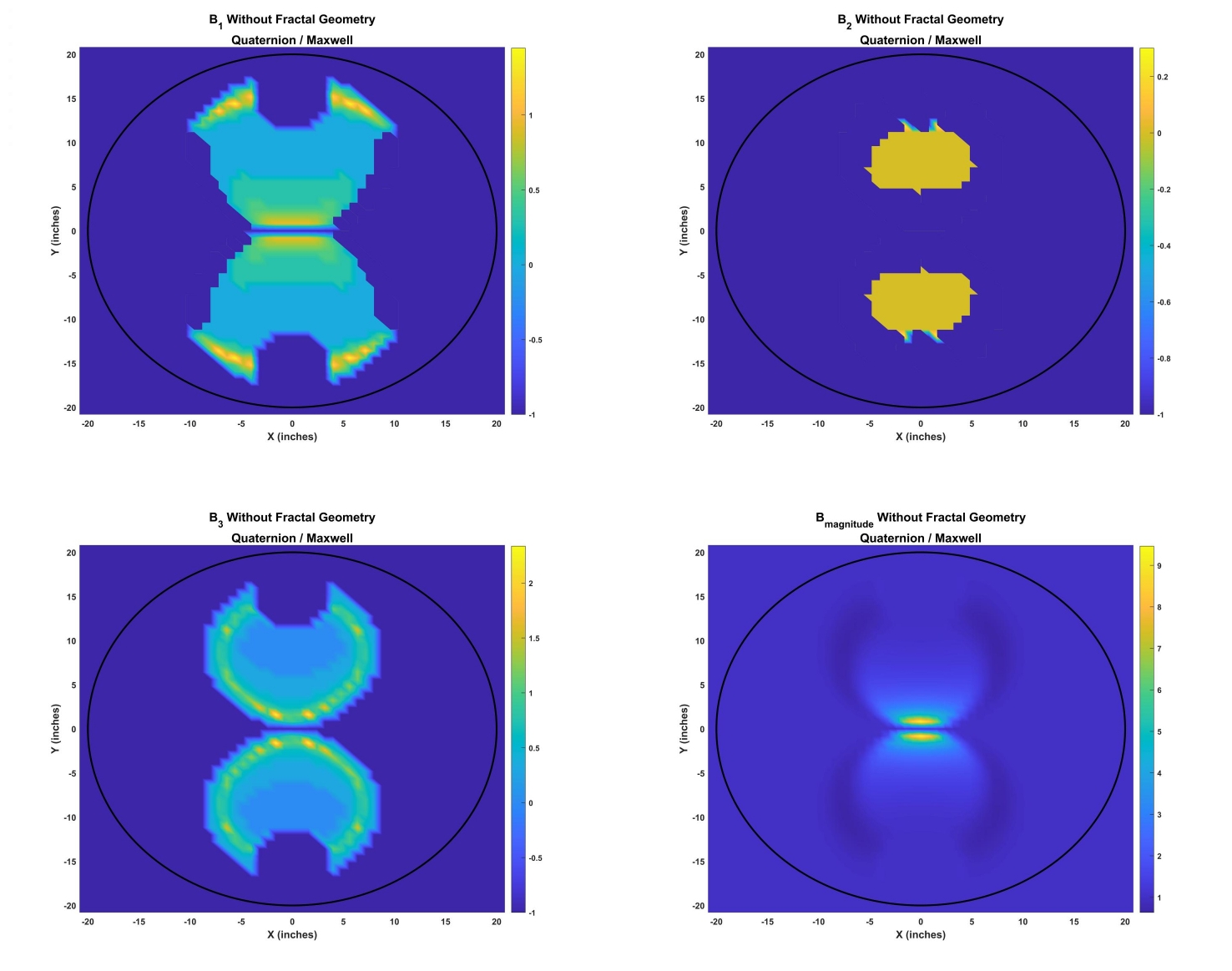}\\
\end{minipage}
    \label{fig:fig13}
    \caption{Moving observer over eddy brake, Study 1 (No Fractal Geometry, No Rotating / Random Wobble) results.}
\end{figure}

\begin{figure}
\centering
\begin{minipage}[b]{0.85\linewidth}
\centering
    \includegraphics[width=\textwidth]{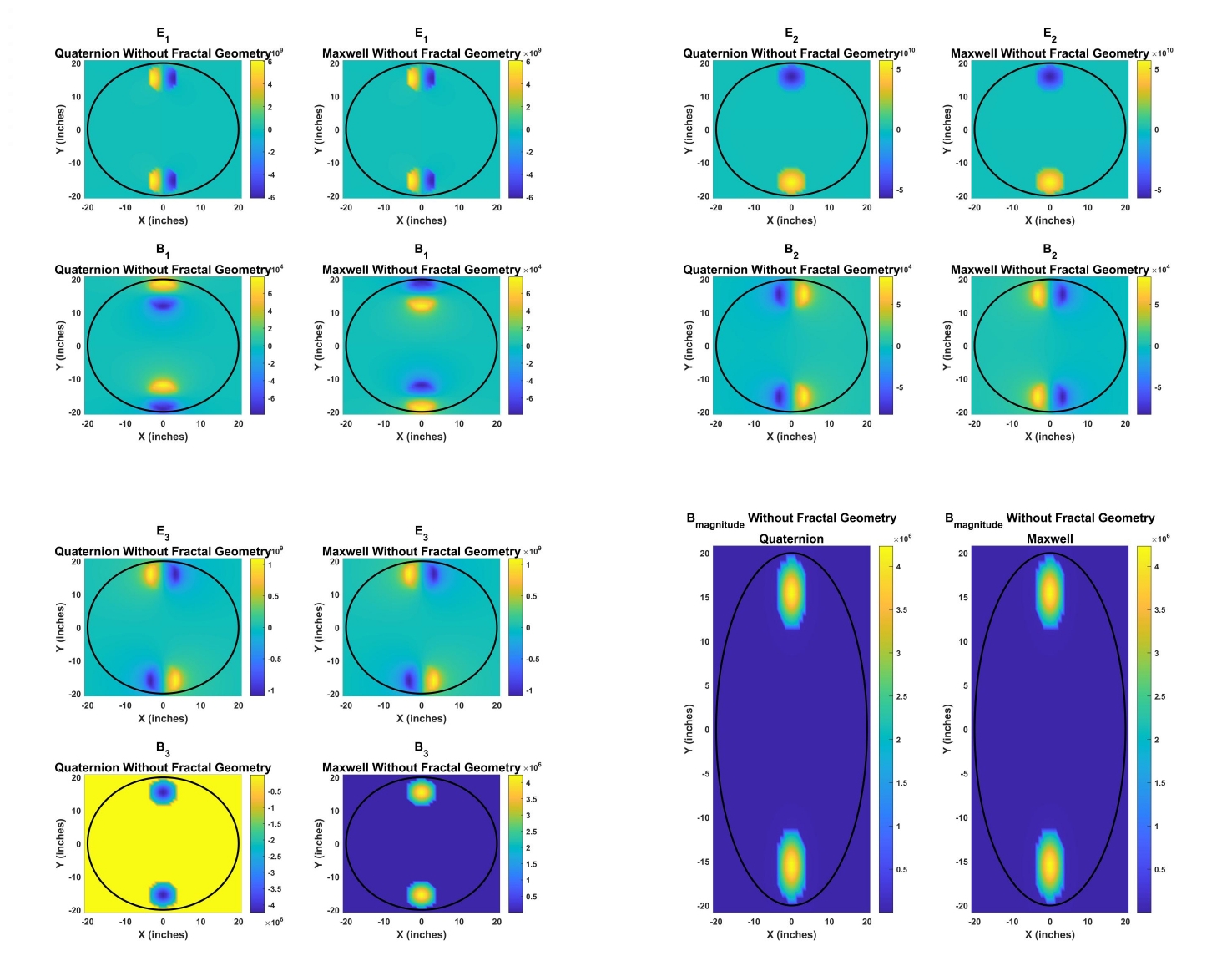}\\
\end{minipage}
\begin{minipage}[b]{0.85\linewidth}
\centering
    \includegraphics[width=\textwidth]{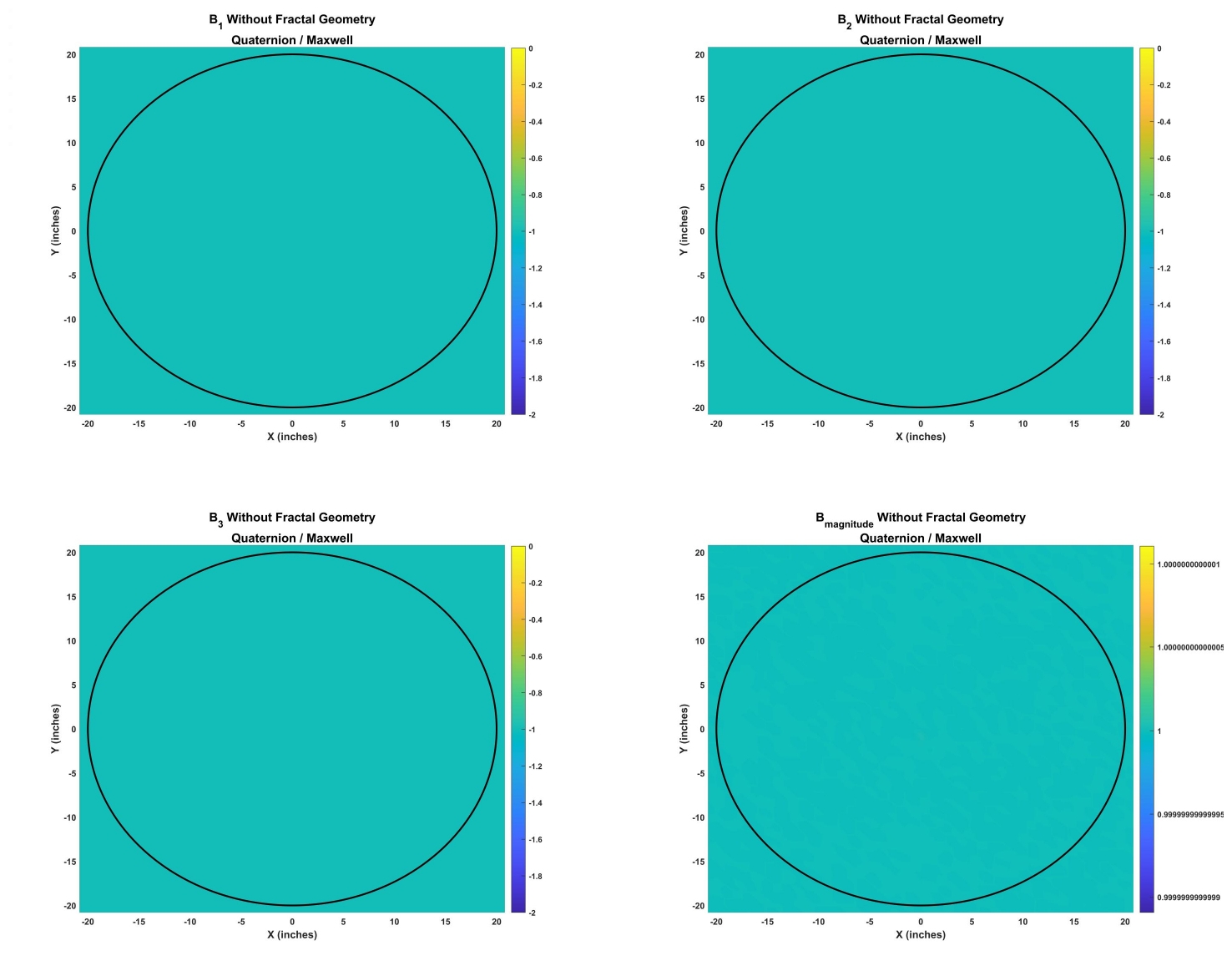}\\
\end{minipage}
    \label{fig:fig14}
    \caption{Moving observer over eddy brake, Study 2 (No Fractal Geometry, Rotating) results.}
\end{figure}

\begin{figure}
\centering
\begin{minipage}[b]{0.85\linewidth}
\centering
    \includegraphics[width=\textwidth]{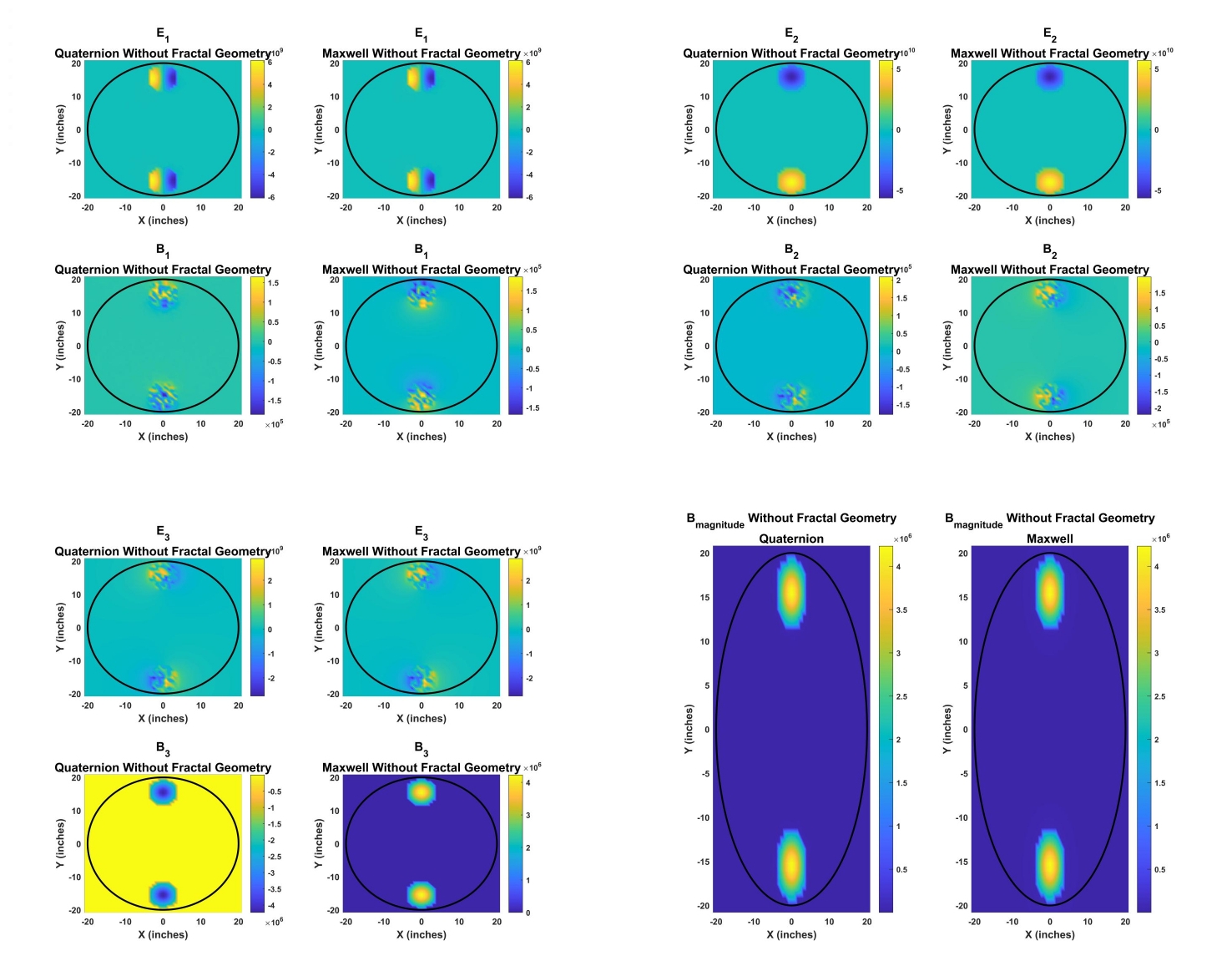}\\
\end{minipage}
\begin{minipage}[b]{0.85\linewidth}
\centering
    \includegraphics[width=\textwidth]{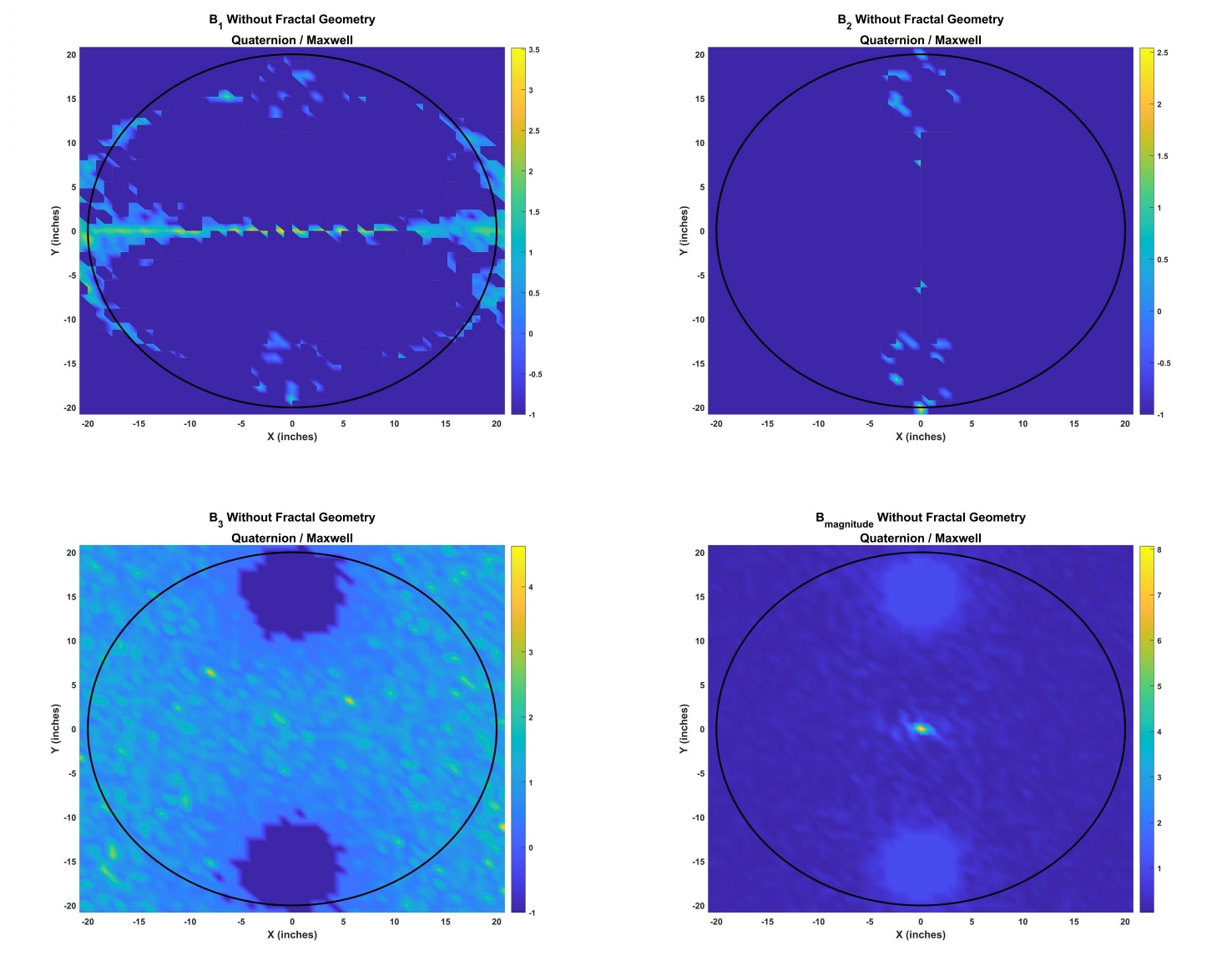}\\
\end{minipage}
    \label{fig:fig15}
    \caption{Moving observer over eddy brake, Study 3 (No Fractal Geometry, Random Wobble) results.}
\end{figure}

\begin{figure}
\centering
\begin{minipage}[b]{0.85\linewidth}
\centering
    \includegraphics[width=\textwidth]{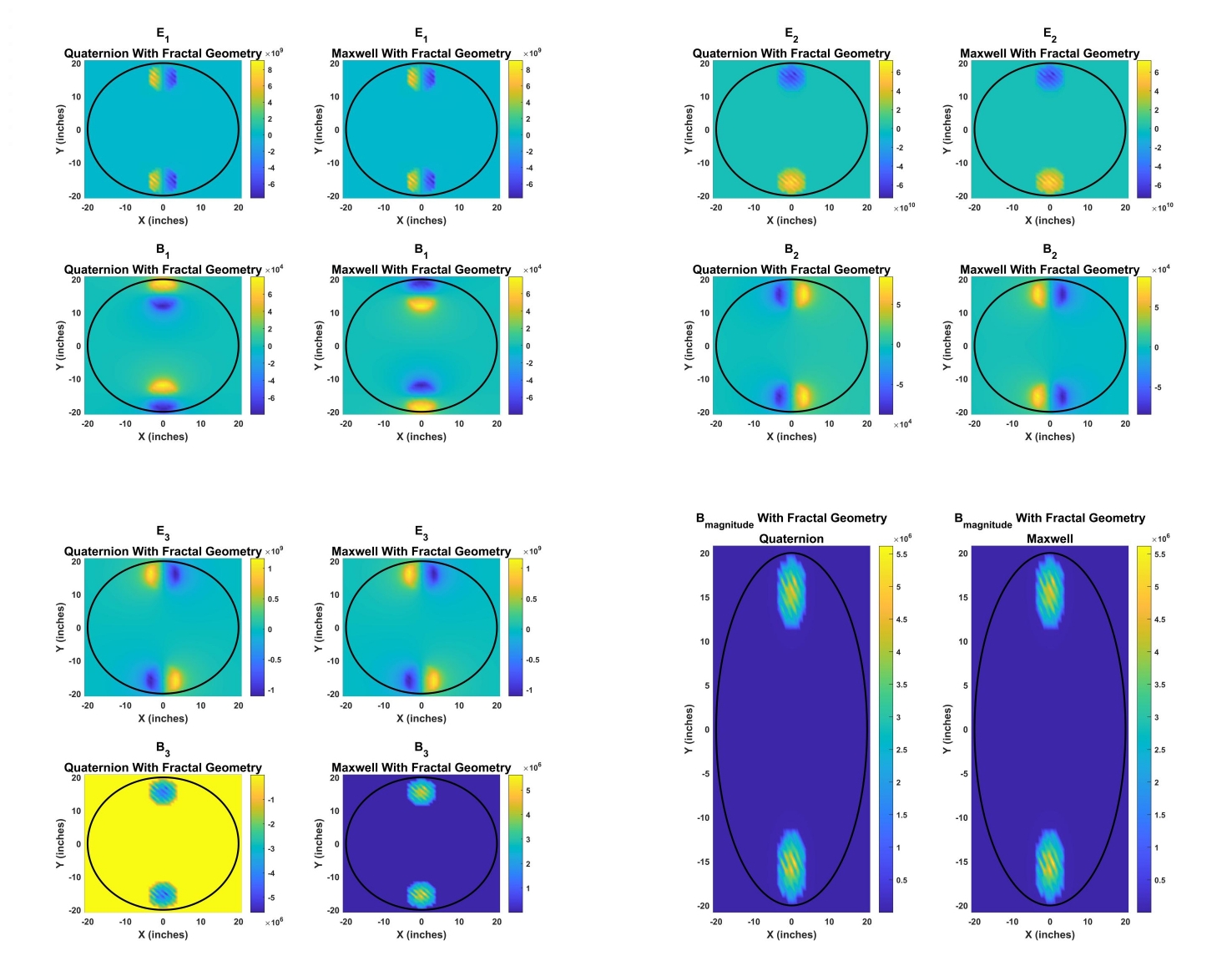}\\
\end{minipage}
\begin{minipage}[b]{0.85\linewidth}
\centering
    \includegraphics[width=\textwidth]{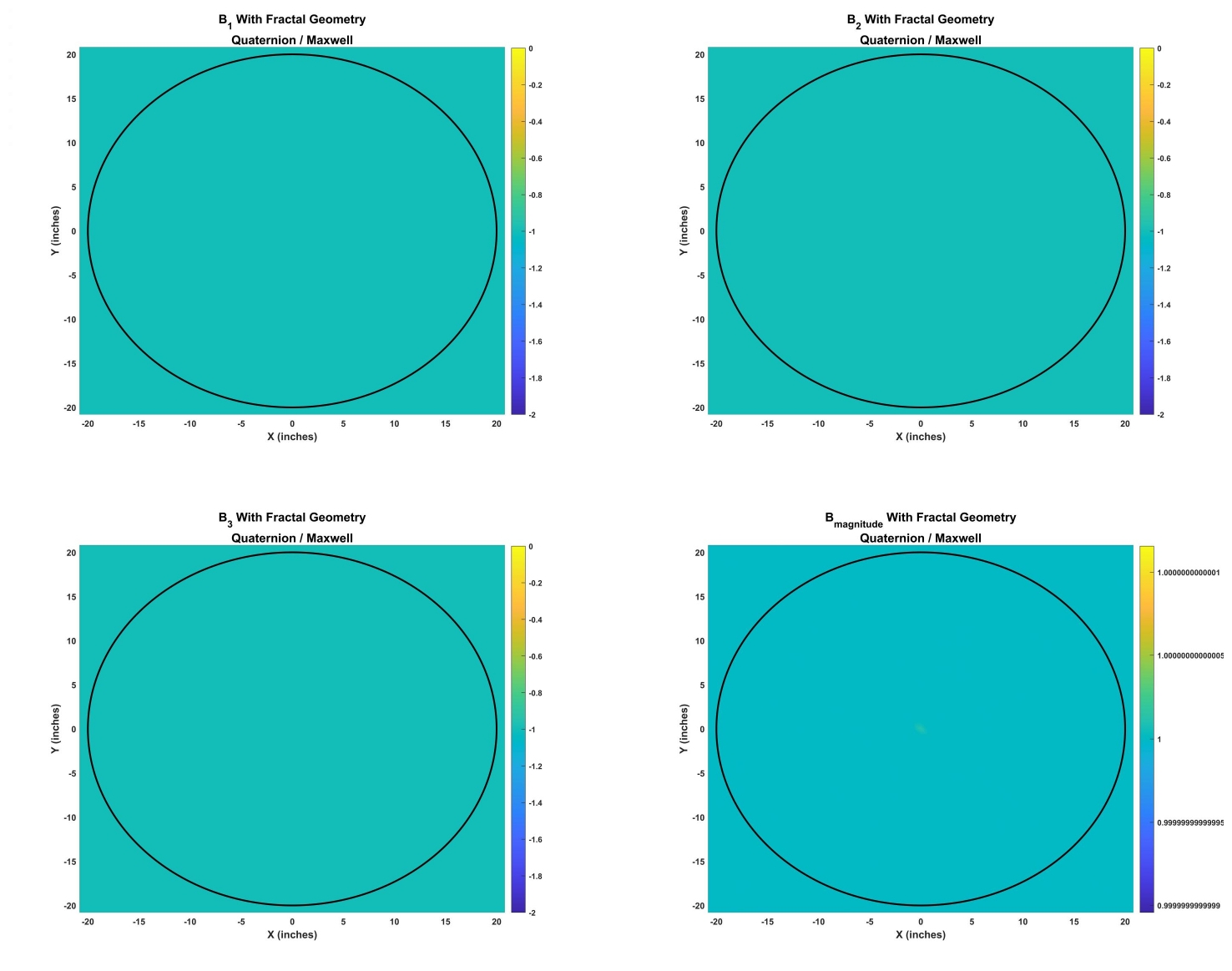}\\
\end{minipage}
    \label{fig:fig16}
    \caption{Moving observer over eddy brake, Study 4 (Fractal Geometry, No Rotating / Random Wobble) results.}
\end{figure}

\begin{figure}
\centering
\begin{minipage}[b]{0.85\linewidth}
\centering
    \includegraphics[width=\textwidth]{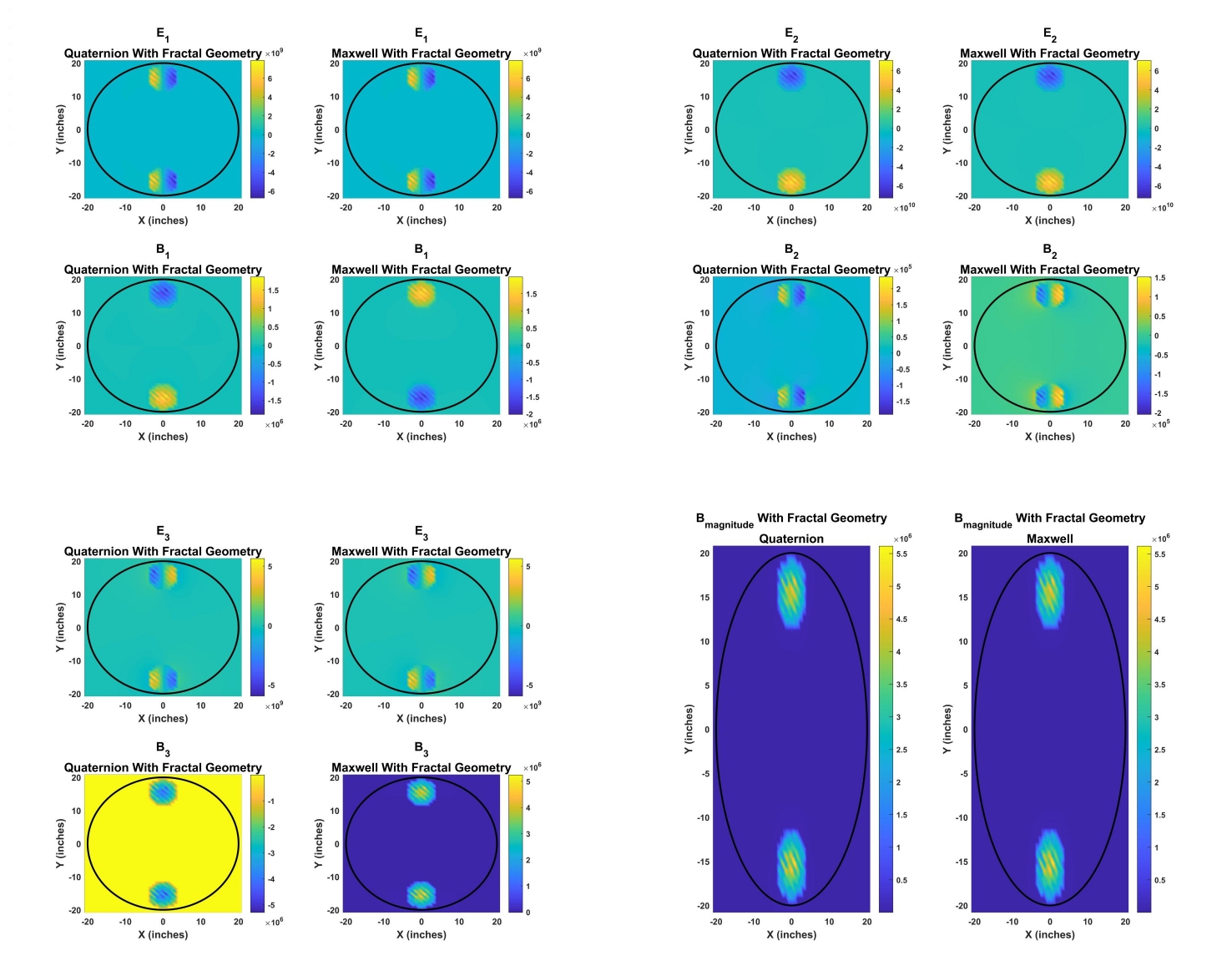}\\
\end{minipage}
\begin{minipage}[b]{0.85\linewidth}
\centering
    \includegraphics[width=\textwidth]{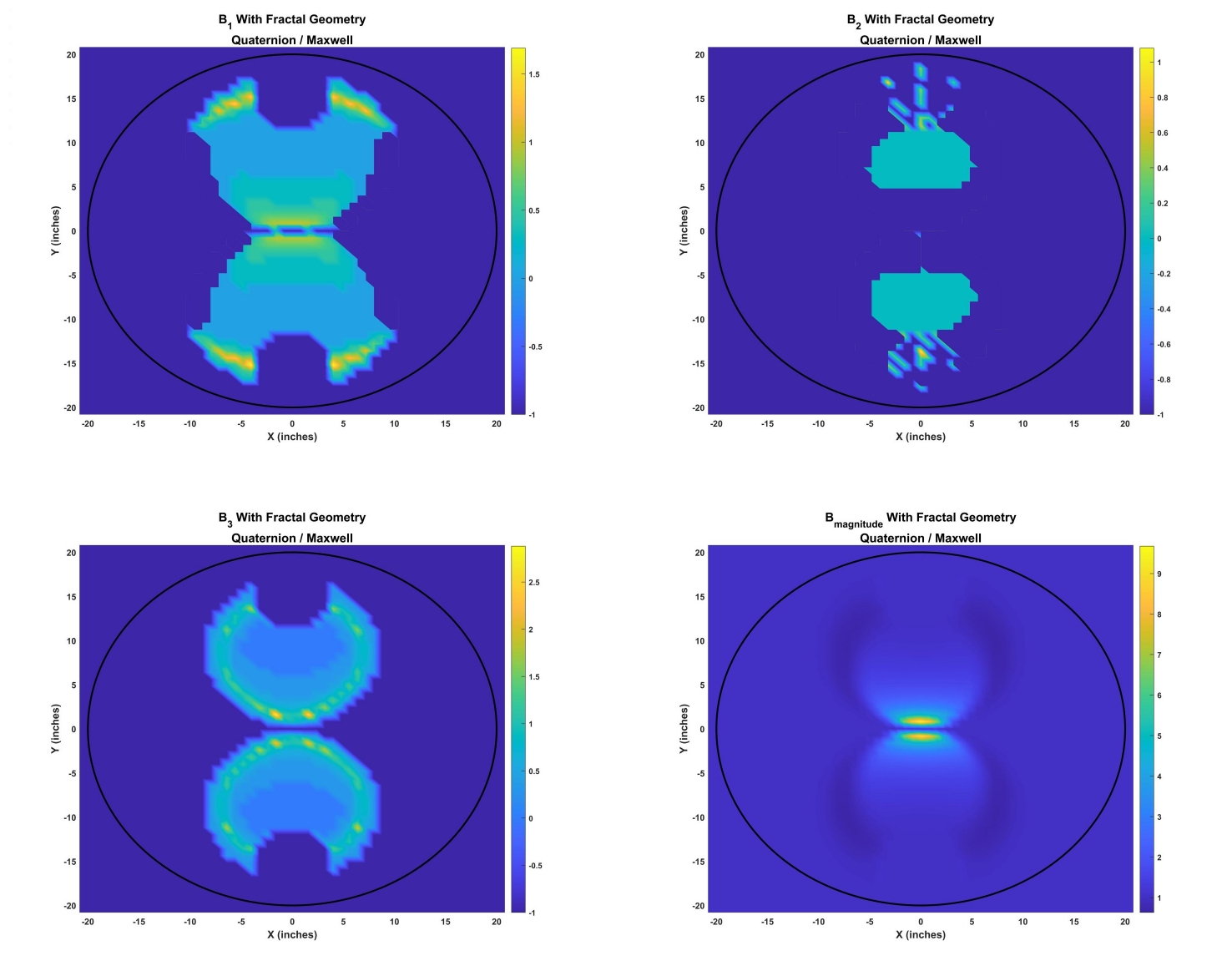}\\
\end{minipage}
    \label{fig:fig17}
    \caption{Moving observer over eddy brake, Study 5 (Fractal Geometry, Rotating) results.}
\end{figure}

\begin{figure}
\centering
\begin{minipage}[b]{0.85\linewidth}
\centering
    \includegraphics[width=\textwidth]{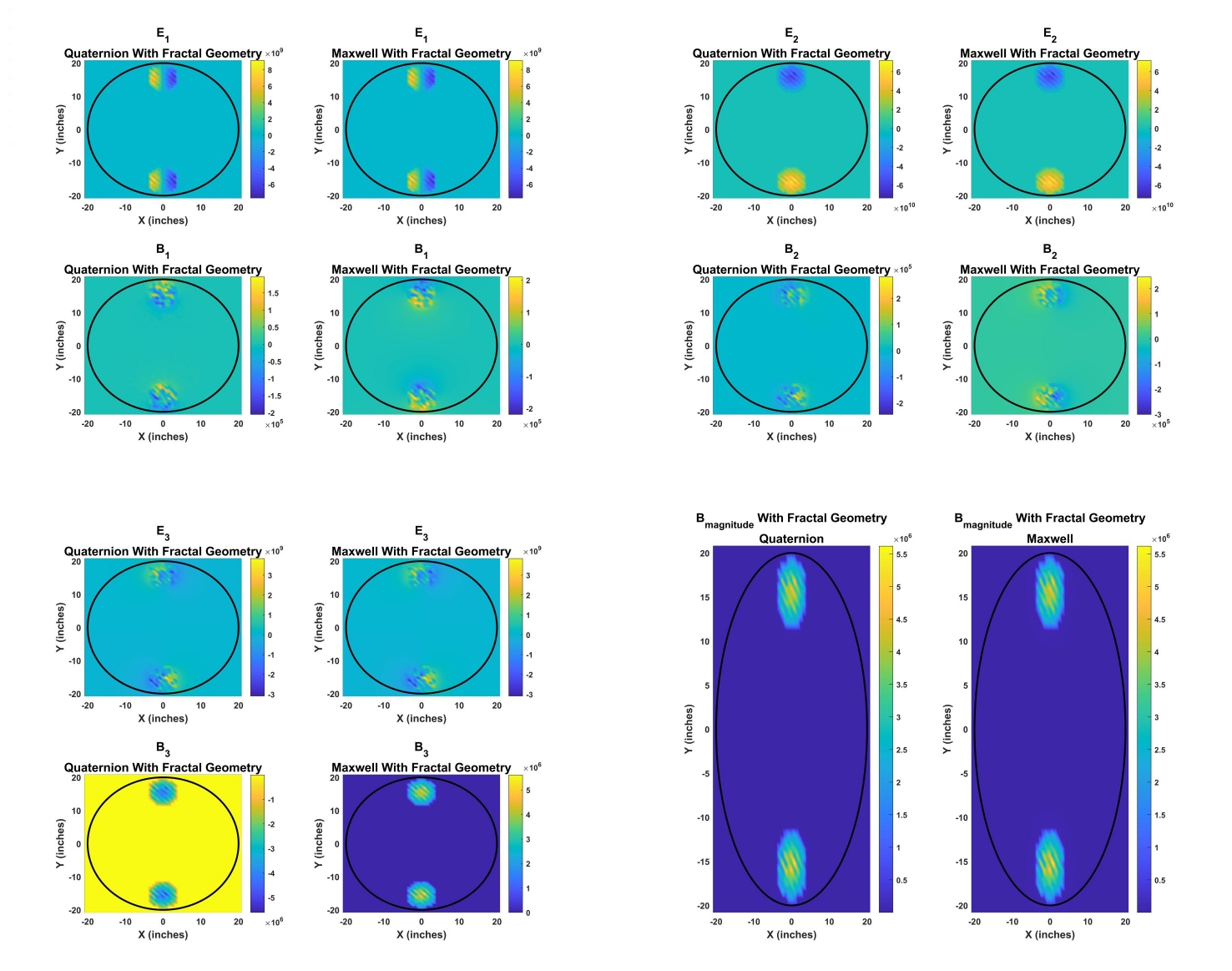}\\
\end{minipage}
\begin{minipage}[b]{0.85\linewidth}
\centering
    \includegraphics[width=\textwidth]{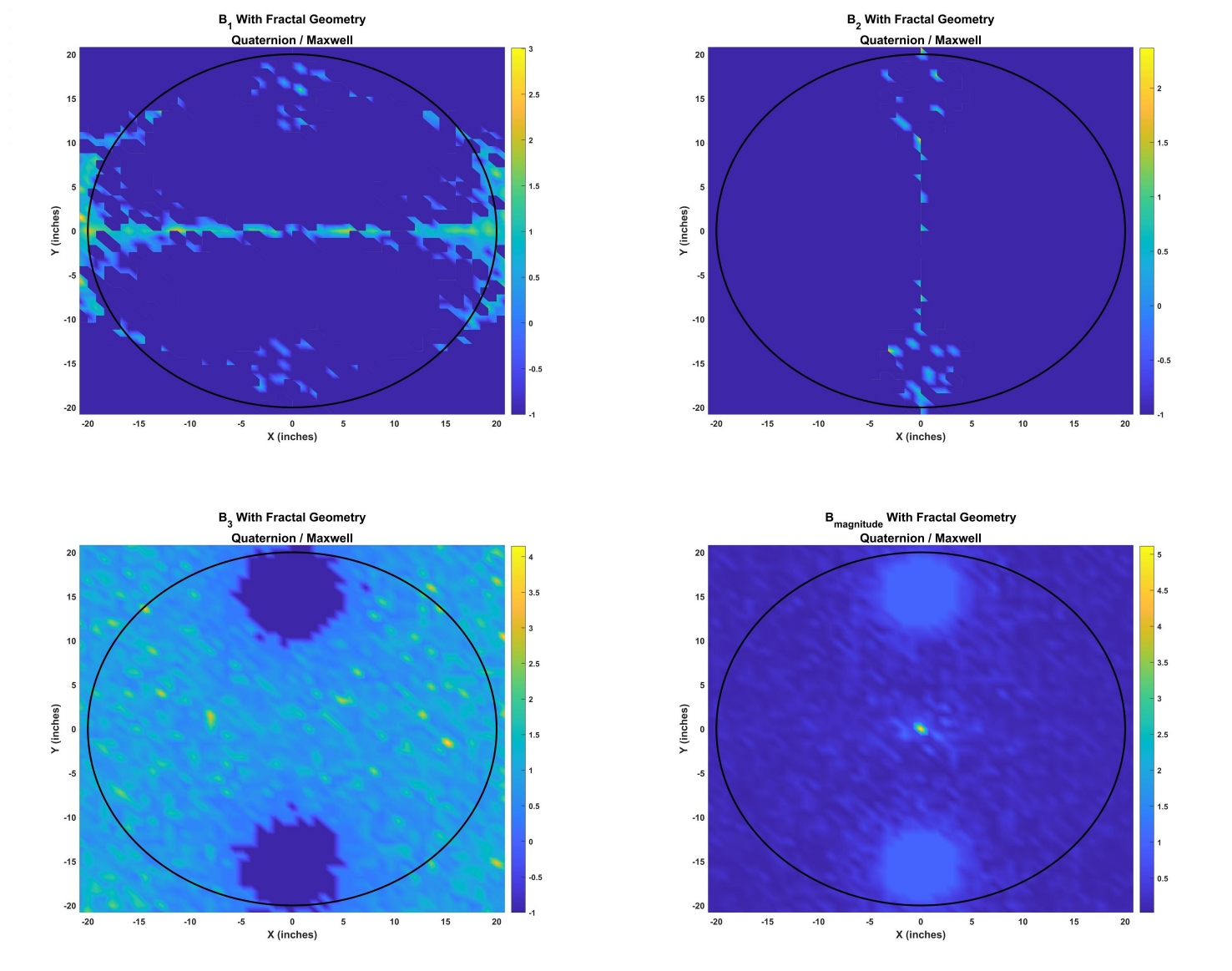}\\
\end{minipage}
    \label{fig:fig18}
    \caption{Moving observer over eddy brake, Study 6 (Fractal Geometry, Random Wobble) results.}
\end{figure}

The six parametric studies are as follows: 
\begin{itemize}
\item Study 1: No Fractal Geometry, No Rotating / Random Wobble (Figure \ref{fig:fig13})
\item Study 2: No Fractal Geometry, Rotating (Figure \ref{fig:fig14})
\item Study 3: No Fractal Geometry, Random Wobble (Figure \ref{fig:fig15})
\item Study 4: Fractal Geometry, No Rotating / Random Wobble (Figure \ref{fig:fig16})
\item Study 5: Fractal Geometry, Rotating (Figure \ref{fig:fig17})
\item Study 6: Fractal Geometry, Random Wobble (Figure \ref{fig:fig18})
\end{itemize}
In Figures \ref{fig:fig13}-\ref{fig:fig18}, the measured electromagnetic and magnetic fields are plotted in the (a) \emph{X}, (b), \emph{Y}, (c) \emph{Z} direction for both Maxwell's and Quaternion; the (d) magnitude of the magnetic field, for both Maxwell's and Quaternion; the ratio of Quaternion over Maxwell's for the magnetic field in the (e) \emph{X}, (f), \emph{Y}, (g) \emph{Z} direction, as well as (h) the absolute magnitude.

\section{Conclusion}

This effort demonstrated how quaternion geometry can better predict the observed magnetic fields, as it overcomes the deficiency with the commonly used Maxwell's equations in separating polar and axial vectors.  The electromagnetic field {\bf{E}} is a polar vector, whereas the magnetic field {\bf{B}} is an axial vector, where the direction of rotation remains the same even after the axial vector is inverted; use of quaternion modeling helps to overcome this error that leads to inverted magnetic results.  The complex quaternion mathematics were solved to a simplified approach, where the rotation matrix was divided by its determinant, leading to identical results.  Parametric studies were analyzed, both of a series of charges, followed by a simulated observer / target, and later an eddy current dynamic electric brake was modeled and the magnetic properties at different positions were compared with traditional Maxwell's analysis.  In all cases, discrepancies are observed, suggesting that the quaternion approach is most accurate for magnetic modeling and analysis of measured results from an observer.  

Future work will address the mathematical augmentation of the geometric algebra model with fractal components. The current work provided a fractal "noise" model that was incorporated into the data. This was done for two purposes: to produce the randomness actually found in nature, and determine if the consequence would produce any instability or anomalous behaviors in the quaternion equations derived for this research effort. There is sufficient reason to expand the electromagnetic field equations to include a fractal component which has been observed in both natural and synthetic environments.

%%%%%%%%%%%%%%%%%%%%%%%%%%%%%%%%%%%%%%%%%%%%%%%%%%%%%%%%%%%%%%%%%%%%%%
%%%%%%%%%%%%%%%%%%%%%%%%%%%%%%%%%%%%%%%%%%%%%%%%%%%%%%%%%%%%%%%%%%%%%%
%%%%%%%%%%%%%%%%%%%%%%%%%%%%%%%%%%%%%%%%%%%%%%%%%%%%%%%%%%%%%%%%%%%%%%
%%%%%%%%%%%%%%%%%%%%%%%%%%%%%%%%%%%%%%%%%%%%%%%%%%%%%%%%%%%%%%%%%%%%%%
%%%%%%%%%%%%%%%%%%%%%%%%%%%%%%%%%%%%%%%%%%%%%%%%%%%%%%%%%%%%%%%%%%%%%%

\clearpage

%%%%%%%%%%%%%%%%%%%%%%% References %%%%%%%%%%%%%%%%%%%%%%%%%

%\section*{References}

%%%%%%%%%%%%%%%%%%%%%%%%%%%%%%%%%%%%%%%%%%%%%%%%%%%%%%%%%%%%%%%%%%%%%%
%\bibliographystyle{unsrt}
%\bibliography{RefFile}

\end{document}